\documentclass[twocolumn,showpacs,preprintnumbers,amsmath,amssymb,superscriptaddress,nofootinbib,english]{revtex4-1}
\usepackage{times,amsmath,amsfonts,amssymb,epstopdf}
\usepackage{graphicx}
\usepackage{dcolumn}
\usepackage{bm}
\usepackage{epsfig}
\usepackage{graphicx}
\usepackage{hyperref}
\usepackage[usenames]{color}
\usepackage{url}
\usepackage[normalem]{ulem}
\usepackage[T1]{fontenc}

\usepackage[usenames]{color}

\def\be{\begin{equation}}
\def\ee{\end{equation}}
\def\ben{\begin{eqnarray}}
\def\een{\end{eqnarray}}
\def\ba{\begin{array}}
\def\ea{\end{array}}

\newcommand{\bq}{\begin{eqnarray}}
\newcommand{\eq}{\end{eqnarray}}
\newcommand{\bes}{\begin{subequations}}
\newcommand{\ees}{\end{subequations}}

\begin{document}
\newcommand{\half}{{\textstyle\frac{1}{2}}}
\allowdisplaybreaks[3]
\def\triangledown{\nabla}
\def\grad3{\hat{\nabla}}
\def\a{\alpha}
\def\b{\beta}
\def\g{\gamma}\def\G{\Gamma}
\def\d{\delta}\def\D{\Delta}
\def\ep{\epsilon}
\def\et{\eta}
\def\z{\zeta}
\def\t{\theta}\def\T{\Theta}
\def\l{\lambda}\def\L{\Lambda}
\def\m{\mu}
\def\f{\phi}\def\F{\Phi}
\def\n{\nu}
\def\r{\rho}
\def\s{\sigma}\def\S{\Sigma}
\def\ta{\tau}
\def\x{\chi}
\def\o{\omega}\def\O{\Omega}
\def\k{\kappa}
\def\pa {\partial}
\def\ov{\over}
\def\br{\\}
\def\ud{\underline}

\def\lcdm{\Lambda{\rm CDM}}
\def\qcdm{{\rm QCDM}}
\def\nloc{R\Box^{-2}R}
\def\msun{M_{\odot}/h}
\def\dw{f(X)}

\newcommand\lsim{\mathrel{\rlap{\lower4pt\hbox{\hskip1pt$\sim$}}
    \raise1pt\hbox{$<$}}}
\newcommand\gsim{\mathrel{\rlap{\lower4pt\hbox{\hskip1pt$\sim$}}
    \raise1pt\hbox{$>$}}}
\newcommand\esim{\mathrel{\rlap{\raise2pt\hbox{\hskip0pt$\sim$}}
    \lower1pt\hbox{$-$}}}
\newcommand{\dpar}[2]{\frac{\partial #1}{\partial #2}}
\newcommand{\sdp}[2]{\frac{\partial ^2 #1}{\partial #2 ^2}}
\newcommand{\dtot}[2]{\frac{d #1}{d #2}}
\newcommand{\sdt}[2]{\frac{d ^2 #1}{d #2 ^2}}    

\title{Nonlinear structure formation in Nonlocal Gravity}

\author{Alexandre Barreira}
\email[Electronic address: ]{a.m.r.barreira@durham.ac.uk}
\affiliation{Institute for Computational Cosmology, Department of Physics, Durham University, Durham DH1 3LE, U.K.}
\affiliation{Institute for Particle Physics Phenomenology, Department of Physics, Durham University, Durham DH1 3LE, U.K.}

\author{Baojiu Li}
\affiliation{Institute for Computational Cosmology, Department of Physics, Durham University, Durham DH1 3LE, U.K.}

\author{Wojciech A. Hellwing}
\affiliation{Institute for Computational Cosmology, Department of Physics, Durham University, Durham DH1 3LE, U.K.}
\affiliation{Interdisciplinary Centre for Mathematical and Computational Modeling (ICM), University of Warsaw, ul. Pawi\'nskiego 5a, Warsaw, Poland}

\author{Carlton M. Baugh}
\affiliation{Institute for Computational Cosmology, Department of Physics, Durham University, Durham DH1 3LE, U.K.}

\author{Silvia Pascoli}
\affiliation{Institute for Particle Physics Phenomenology, Department of Physics, Durham University, Durham DH1 3LE, U.K.}

\preprint{IPPP/14/ 70 DCPT/14/ 140}

\begin{abstract}

We study the nonlinear growth of structure in nonlocal gravity models with the aid of N-body simulation and the spherical collapse and halo models. We focus on a model in which the inverse-squared of the d'Alembertian operator acts on the Ricci scalar in the action. For fixed cosmological parameters, this model differs from $\lcdm$ by having a lower late-time expansion rate and an enhanced and time-dependent gravitational strength ($\sim 6\%$ larger today). Compared to $\lcdm$ today, in the nonlocal model, massive haloes are slightly more abundant (by $\sim 10\%$ at $M \sim 10^{14} \msun$) and concentrated ($\approx 8\%$ enhancement over a range of mass scales), but their linear bias remains almost unchanged. We find that the Sheth-Tormen formalism describes the mass function and halo bias very well, with little need for recalibration of free parameters. The fitting of the halo concentrations is however essential to ensure the good performance of the halo model on small scales. For $k \gtrsim 1 h/{\rm Mpc}$, the amplitude of the nonlinear matter and velocity divergence power spectra exhibits a modest enhancement of $\sim 12\%$ to $15\%$, compared to $\lcdm$ today. This suggests that this model might only be distinguishable from $\lcdm$ by future observational missions. We point out that the absence of a screening mechanism may lead to tensions with Solar System tests due to local time variations of the gravitational strength, although this is subject to assumptions about the local time evolution of background averaged quantities.

\end{abstract} 
%\pacs{98.80.Cq}
\maketitle

\section{Introduction}

It has been almost a century since Einstein proposed his theory of General Relativity (GR) which is still considered one of the main pillars of modern physics. The outstanding success of GR comes mostly from its ability to pass a number of stringent tests of gravity performed in the Solar System \cite{Will:2014kxa}. When applied on cosmological scales, however, GR seems to lose some of its appeal as it requires the presence of some unknown form of {\it dark energy} in order to explain the observed accelerated expansion of the Universe. The simplest candidate for dark energy is a cosmological constant, $\Lambda$, but the value of $\Lambda$ required to explain the observations lacks theoretical support. This has provided motivation for the proposal of alternative gravity models which attempt to reproduce cosmic acceleration without postulating the existence of dark energy. Furthermore, the fact that the laws of gravity have never been tested directly on scales larger than the Solar System justifies the exploration of such modifications to GR on cosmological scales. By understanding better the various types of observational signatures that different {\it modified gravity} models can leave on cosmological observables, one can improve the chance of identifying any departures from GR, or alternatively, extend the model's observational success into a whole new regime. Currently, the study of modified gravity models is one of the most active areas of research in both theoretical and observational cosmology \cite{Clifton:2011jh, Huterer:2013xky, Jain:2013wgs, Joyce:2014kja}.

Here, we focus on a class of model that has attracted much attention recently, which is known as {\it nonlocal gravity} \cite{Woodard:2014iga}. In these models, the modifications to gravity arise via the addition of nonlocal terms (i.e.~which depend on more than one point in spacetime) to the Einstein field equations. These terms typically involve the inverse of the d'Alembertian operator, $\Box^{-1}$, acting on curvature tensors. To ensure causality, such terms must be defined with the aid of retarded Green functions (or propagators). However, it is well known that such retarded operators cannot be derived from standard action variational principles (see e.g.~Sec.~2 of Ref.~\cite{Woodard:2014iga} for a discussion). One way around this is to specify the model in terms of its equations of motion and not in terms of its action. One may still consider a nonlocal action to derive a set of causal equations of motion, so long as in the end one replaces, by hand, all of the resulting operators by their retarded versions. Both of these approaches, however, imply that nonlocal models of gravity must be taken as purely phenomenological and should not be interpreted as fundamental theories. In general, one assumes that there is an unknown fundamental (local) quantum field theory of gravity, and the nonlocal model represents only an effective way of capturing the physics of the fundamental theory in some appropriate limit.

It was in the above spirit that Ref.~\cite{Deser:2007jk} proposed a popular nonlocal model of gravity capable of explaining cosmic acceleration. In this model, which has been extensively studied (see e.g. Refs.~\cite{Deffayet:2009ca, Deser:2013uya, Woodard:2014iga, Nojiri:2007uq, Jhingan:2008ym, Koivisto:2008xfa, Koivisto:2008dh, 2012PhRvD..85d4002E, 2013CQGra..30c5002E, 2013PhRvD..87b4003P, Dodelson:2013sma} and references therein), one adds the term $R f\left(\Box^{-1}R\right)$ to the Einstein-Hilbert action, where $R$ is the Ricci scalar and $f$ is a free function. As described in Ref.~\cite{Woodard:2014iga}, the function $f$ can be constructed in such a way that it takes on different values on the cosmological background and inside gravitationally bound systems. In particular, at the background level, $f$ can be tuned to reproduce $\lcdm$-like expansion histories, but inside regions like the Solar System, one can assume that $f$ vanishes, thus recovering GR completely. This model, however, seems to run into tension with data sensitive to the growth rate of structure on large scales \cite{2013PhRvD..87b4003P, Dodelson:2013sma}. 

More recently, nonlocal terms have also been used to construct theories of massive gravity. An example of this is obtained by adding directly to the Einstein field equations a term like $m^2\left(g_{\mu\nu}\Box^{-1}R\right)^T$ \cite{Maggiore:2013mea, Foffa:2013vma, Kehagias:2014sda, Nesseris:2014mea}, where $m$ is a mass scale and $^T$ means the extraction of the transverse part (see also Refs.~\cite{Jaccard:2013gla, Foffa:2013sma, Modesto:2013jea, Ferreira:2013tqn} for models in which $\Box^{-1}$ acts on the Einstein and Ricci tensors). This model has no $\lcdm$ limit for the background evolution, but it can still match the current background expansion and growth rate of structure data with a similar goodness-of-fit \cite{Nesseris:2014mea}. Furthermore, Ref.~\cite{Kehagias:2014sda} has investigated spherically symmetric static solutions in this model, concluding that it does not suffer from instabilities that usually plague theories of massive gravity. A similar model was proposed by Ref.~\cite{Maggiore:2014sia}, which is characterized by a term $\propto m^2R\Box^{-2}R$ in the action (see Eq.~(\ref{eq:action})). Reference \cite{Dirian:2014ara} showed that this model can reproduce current type Ia Supernovae (SNIa) data, although it also has no $\lcdm$ limit for the background expansion. The time evolution of linear matter density fluctuations in this model also differs from that in $\lcdm$, but the work of Ref.~\cite{Dirian:2014ara} suggests that the differences between these two models are small enough to be only potentially distinguishable by future observational missions.

Here, we extend the previous work done for the model of Refs.~\cite{Maggiore:2014sia, Dirian:2014ara} by examining its predictions in the nonlinear regime of structure formation. We achieve this by running a set of  N-body simulations, which we use to analyse the model predictions for the nonlinear matter and velocity divergence power spectra, and also halo properties such as their abundance, bias and concentration. To the best of our knowledge this is the first time N-body simulations have been used to study the nonlinear regime of structure formation in nonlocal gravity cosmologies. N-body simulations are however computationally expensive to run. To overcome this, it is common to try to devise semi-analytical formulae that, motivated by simple physical assumptions, aim to reproduce the results from the simulations with the calibration of free parameters kept to a minimum. A popular example is given by the Sheth-Tormen halo mass function \cite{Sheth:1999mn, Sheth:1999su, Sheth:2001dp} and its use in the halo model approach for the nonlinear matter power spectrum \cite{Cooray:2002dia}. The performance of the halo model is well understood within $\lcdm$, but less so in alternative gravity scenarios. In particular, the free parameters in these formulae might need substantial recalibration in models that differ significantly from $\lcdm$ (see e.g.~Ref.~\cite{2014JCAP...04..029B}). One of our goals is to assess the performance of these analytical formulae in nonlocal gravity models.

This paper is organized as follows. In Sec.~\ref{sec:model} we present the model and layout the equations relevant for the background evolution. We also derive the equations of motion for spherically symmetric configurations under the quasi-static and weak-field approximations. In Sec.~\ref{sec:halo-model-formulae} we present the formulae relevant for the calculation of the nonlinear matter power spectrum in the halo model formalism. In particular, we describe the Sheth-Tormen expressions for the halo mass function and bias, and define the Navarro-Frenk-White (NFW) halo concentration parameter. We also present the equations relevant for the linear evolution and spherical collapse of matter overdensities. Our results are presented in Sec.~\ref{sec:results}, where we discuss the results from the N-body simulations and compare them with the predictions from the analytical formulae. We also comment on the role that Solar System tests of gravity could play in setting the observational viability of this model. We summarize our findings in Sec.~\ref{sec:summary}.

In this paper, we work with the metric signature $\left(+,-,-,-\right)$ and use units in which the speed of light $c = 1$. Latin indices run over $1,2,3$ and Greek indices run over $0,1,2,3$. We use $\kappa = 8\pi G = 1/M_{\rm Pl}^2$ interchangeably, where $M_{\rm Pl}$ is the reduced Planck mass and $G$ is Newton's constant.

\section{The $\nloc$ nonlocal gravity model}\label{sec:model}

\subsection{Action and field equations}

We consider the nonlocal gravity model of Refs.~\cite{Maggiore:2014sia, Dirian:2014ara}, whose action is given by

\bq\label{eq:action}
A = \frac{1}{2\kappa}\int {\rm d}x^4\sqrt{-g}\left[R - \frac{m^2}{6}R\Box^{-2}R - \mathcal{L}_m\right],
\eq
where $g$ is the determinant of the metric $g_{\mu\nu}$, $\mathcal{L}_m$ is the Lagrangian density of the matter fluid, $R$ is the Ricci scalar and $\Box = \nabla^\mu\nabla_\mu$ is the d'Alembertian operator. To facilitate the derivation of the field equations, and to solve them afterwards, it is convenient to introduce two auxiliary scalar fields defined as

\bq
\label{eq:u}U &=& - \Box^{-1} R ,\ \ \ \ \ \ \\
\label{eq:s}S &=& -\Box^{-1} U = \Box^{-2}R.
\eq
The solutions to Eqs.~(\ref{eq:u}) and (\ref{eq:s}) can be obtained by evaluating the integrals

\bq
\label{eq:intU}U &\equiv& -\Box^{-1}R  \\ \nonumber
&=& U_{\rm hom}(x) - \int {\rm d}^4y \sqrt{-g(y)}G(x,y)R(y), \\
\label{eq:intS}S &\equiv& -\Box^{-1}U \\ \nonumber
&=& S_{\rm hom}(x) - \int {\rm d}^4y \sqrt{-g(y)}G(x,y)U(y),
\eq
where $U_{\rm hom}$ and $S_{\rm hom}$ are any solutions of the homogeneous equations $\Box U = 0$ and $\Box S = 0$, respectively, and $G(x, y)$ is any Green function of $\Box$. The choice of the homogeneous solutions and of the Green function specify the meaning of the operator $\Box^{-1}$. To ensure causality, one should use the retarded version of the Green function, i.e., the solutions of $U$ (or $S$) should only be affected by the values of $R$ (or $U$) that lie in its past light-cone. The homogeneous solutions can be set to any value, which is typically zero, without any loss of generality. In principle, the model predictions can be obtained by solving Eqs.~(\ref{eq:intU}) and (\ref{eq:intS}). However, it is convenient to use the fields $U$ and $S$ to cast the nonlocal action of Eq.~(\ref{eq:action}) in the form of a local scalar-tensor theory \cite{Nojiri:2007uq, Capozziello:2008gu, Koshelev:2008ie} as 

\bq\label{eq:action-local}
A &&= \frac{1}{2\kappa}\int {\rm d}x^4\sqrt{-g}\left[R - \frac{m^2}{6}RS - \xi_1\left(\Box U + R\right) \right. \nonumber \\
&&\ \ \ \ \ \ \ \ \ \ \ \ \  \ \ \ \ \ \ \ \ \ \ \ \ \ \ \ \ \ \ \ \left. - \xi_2\left(\Box S + U\right) - \mathcal{L}_m\right],
\eq
where $\xi_1$ and $\xi_2$ are Lagrange multipliers. The field equations can then be written as

\bq
&&\label{eq:fe1}G_{\mu\nu} - \frac{m^2}{6}K_{\mu\nu} = \kappa T_{\mu\nu}, \\
&&\label{eq:fe2}\Box U = -R, \\
&&\label{eq:fe3}\Box S = -U,
\eq
with
\bq
\label{eq:fe4}K_{\mu\nu} \equiv 2SG_{\mu\nu} - 2\nabla_\mu\nabla_\nu S - 2\nabla_{(\mu}S\nabla_{\nu)}U \nonumber \\ 
+ \left(2\Box S + \nabla_\alpha S\nabla^\alpha U - \frac{U^2}{2}\right)g_{\mu\nu},
\eq
and where $T^{\mu\nu} = \left(2/\sqrt{-g}\right)\delta\left(\mathcal{L}_m\sqrt{-g}\right)/\delta g_{\mu\nu}$ is the energy-momentum tensor of the matter fluid. The use of the scalar fields $U$ and $S$ therefore allows one to obtain the solutions by solving a set of coupled differential equations, instead of the more intricate integral equations associated with the inversion of a differential operator. These two formulations are, however, not equivalent as explained with detail in many recent papers (see e.g.~Refs.\cite{Koshelev:2008ie, Koivisto:2009jn, Barvinsky:2011rk, Deser:2013uya, Maggiore:2013mea, Foffa:2013sma, Foffa:2013vma}): Eqs.~(\ref{eq:fe1}), (\ref{eq:fe2}) and (\ref{eq:fe3}) admit solutions that are not solutions of the original nonlocal problem. For instance, if $U^*$ is a solution of Eq.~(\ref{eq:fe2}), then $U^* + U_{\rm hom}$ is also a solution for any $U_{\rm hom}$, since $\Box U_{\rm hom} = 0$ (the same applies for the field $S$ and Eq.~(\ref{eq:fe3})). If one wishes the differential equations (\ref{eq:fe1}), (\ref{eq:fe2}) and (\ref{eq:fe3}) to describe the nonlocal model, then one must solve them with the one and only choice of initial conditions that is compatible with the choice of homogeneous solutions in Eqs.~(\ref{eq:intU}) and (\ref{eq:intS}). All other initial conditions lead to spurious solutions and should not be considered.

\subsection{Background equations}

Throughout, we always work with the perturbed Friedmann-Roberston-Walker (FRW) line element in the Newtonian gauge,

\bq\label{eq:ds}
{\rm d}s^2 = \left(1 + 2\Psi\right){\rm d}t^2 - a(t)^2\left(1 - 2\Phi\right)\gamma_{ij}{\rm d}x^i{\rm d}x^j,
\eq
where $a = 1/(1+z)$ is the cosmic scale factor ($z$ is the redshift) and the gravitational potentials $\Phi$, $\Psi$ are assumed to be functions of time and space. $\gamma_{ij}$ is the spatial sector of the metric, which is taken here to be flat.

At the level of the cosmological background ($\Phi = \Psi = 0$), the two Friedmann equations can be written as

\bq
\label{eq:friedmann1} 3H^2 &=& \kappa\bar{\rho}_m + \kappa\bar{\rho}_{de}\\
\label{eq:friedmann2} -2\dot{H} - 3H^2 &=& \kappa\bar{p}_m + \kappa\bar{p}_{de},
\eq
where we have encapsulated the effects of the nonlocal term into an effective background "dark energy" density, $\bar{\rho}_{de}$, and pressure $\bar{p}_{de}$, which are given, respectively, by

\bq
\label{eq:rhode}\kappa\bar{\rho}_{de} &=& \frac{m^2}{6}\left[6\bar{S}H^2 + 6H\dot{\bar{S}} - \dot{\bar{U}}\dot{\bar{S}} - \frac{\bar{U}^2}{2}\right], \\
\label{eq:pde} \kappa\bar{p}_{de} & = & - \frac{m^2}{6}\left[2\bar{S}\left(2\dot{H} + 3H^2\right) + \ddot{\bar{S}} \right.\\ \nonumber
&& \ \ \ \ \ \ \ \ \ \ \ \left.+ 4H\dot{\bar{S}} + \dot{\bar{U}}\dot{\bar{S}} - \frac{\bar{U}^2}{2}\right].
\eq
Additionally, Eqs.~(\ref{eq:fe2}) and (\ref{eq:fe3}) yield

\bq
\label{eq:ubg}\ddot{\bar{U}} + 3H\dot{\bar{U}} &=& 6\left(\dot{H} + 2H^2\right), \\
\label{eq:sbg}\ddot{\bar{S}} + 3H\dot{\bar{S}} &=& -\bar{U}. \\
\eq
In the above equations, a dot denotes a partial derivative w.r.t. physical time, $t$, an overbar indicates that we are considering only the background average and $H = \dot{a}/a$ is the Hubble expansion rate.

The background evolution in the $\nloc$ model has to be obtained numerically. The differential equations are evolved starting from deep into the radiation dominated era ($z = 10^6$) with initial conditions for the auxiliary fields $\bar{U} = \dot{\bar{U}} = \bar{S} = \dot{\bar{S}} = 0$. Note that, in the radiation era, the Ricci scalar vanishes ($\bar{R} = 6\dot{H} + 12H^2 = 0$). Hence, from Eqs.~(\ref{eq:intU}) and (\ref{eq:intS}) one sees that these initial conditions are indeed compatible with the choice $U_{\rm hom} = S_{\rm hom} = 0$. The value of the parameter $m$ is determined by a trial-and-error scheme to yield the value of $\bar{\rho}_{de0}$ that makes the Universe spatially flat,  i.e., $\bar{\rho}_{r0} + \bar{\rho}_{m0} + \bar{\rho}_{de0} = \bar{\rho}_{c0} \equiv 3H_0^2/\kappa$, where the subscripts $_r$, $_m$ refer to radiation and matter, respectively, the subscript $_0$ denotes present-day values, and $H_0 = 100h {\rm km/s/Mpc}$ is the present-day Hubble rate.

\subsection{Spherically symmetric nonlinear equations}

By assuming that the potentials $\Phi$ and $\Psi$ are spherically symmetric, one can write the $(0,0)$ and $(r,r)$ components of Eq.~(\ref{eq:fe1}), and Eqs.~(\ref{eq:fe2}) and (\ref{eq:fe3}), respectively, as

\begin{widetext}
\bq
\label{eq:sph1}&&\frac{2}{r^2}\left(r^2\Phi,_r\right),_r  - \frac{m^2}{6}\left[6SH^2 + \frac{4S}{r^2}\left(r^2\Phi,_r\right),_r - \frac{2}{r^2}\left(r^2S,_r\right),_r + 2S,_r\Phi,_r  - S,_rU,_r - \frac{U^2}{2}\right] = \kappa\bar{\rho}_m\delta a^2, \\
\label{eq:sph2}&&\frac{2}{r}\left(\Phi,_r - \Psi,_r\right) - \frac{m^2}{6}\left[4S\dot{H}a^2 + 6SH^2a^2 + \frac{4S}{r}\left(\Phi,_r - \Psi,_r\right) + 4S,_r\Phi,_r - 2S,_r\Psi,_r - 4\frac{S,_r}{r} + 2S,_rU,_r - \frac{U^2}{2}\right] = 0, \\
\label{eq:sph3}&&\frac{1}{r^2}\left(r^2U,_r\right),_r + U,_r\left(\Psi,_r - \Phi,_r\right)=   2\frac{1}{r^2}\left(r^2\Psi,_r\right),_r - 4\frac{1}{r^2}\left(r^2\Phi,_r\right),_r,  \\
\label{eq:sph4}&&\frac{1}{r^2}\left(r^2S,_r\right),_r + S,_r\left(\Psi,_r - \Phi,_r\right) = U,
\eq
\end{widetext}
where $,_r$ denotes a partial derivative w.r.t. the comoving radial coordinate $r$. When writing Eqs.~(\ref{eq:sph1})-(\ref{eq:sph4}), we have already employed the following simplifying assumptions:

\begin{enumerate}

\item We have assumed the so-called quasi-static limit, under which one neglects the time derivatives of perturbed quantities, e.g., $\dot{S} = \dot{\bar{S}} + \dot{\delta S} \approx \dot{\bar{S}}$, where $\delta S$  is the perturbed part of the auxiliary field;

\item We have also employed the so-called weak-field limit, which accounts for neglecting terms that involve $\Phi$ and $\Psi$, and their first spatial derivatives, over those that involve their second spatial derivatives. For example, $\left(1 - 2\Phi\right)\Phi,_{rr} \approx \Phi,_{rr}$ and $\Phi,_{r}\Phi,_{r} \ll \Phi,_{rr}$. 

\end{enumerate}
The above equations still contain terms with $\Psi,_r$ and $\Phi,_r$, because these terms contain the fields $U$ and $S$, and up to now, we have not discussed the validity of applying these approximations to the auxiliary fields. However:

\begin{enumerate}

\item Equation (\ref{eq:sph3}) tells us that the $U$ field is of the same order as the scalar potentials, $U \sim \Phi, \Psi$. Consequently, the above approximations also hold for $U$;

\item Equation (\ref{eq:sph4}) tells us that $S,_{rr} \sim \Phi,\Psi$, which means we can also neglect all terms containing $S$, $S,_r$ and $S,_{rr}$.

\end{enumerate}
Under these considerations, the above equations simplify drastically. In particular, the only equation that remains relevant for the study of the spherical collapse of matter overdensities is Eq.~(\ref{eq:sph1}), which can be written as:

\bq
\label{eq:modpoisson}\frac{1}{r^2}\left(r^2\Phi,_r\right),_r = 4\pi G_{\rm eff} \bar{\rho}_m \delta a^2,
\eq
where 

\bq
\label{eq:geff}G_{\rm eff} = G \left[1 - \frac{m^2\bar{S}}{3}\right]^{-1}.
\eq
Equation~(\ref{eq:modpoisson}) is the same as in standard gravity, but with Newton's constant replaced by the time-dependent gravitational strength, $G_{\rm eff}$. This time dependence follows directly from the term $2SG_{\mu\nu}$ in the field equations, Eq.~(\ref{eq:fe1}), which in turn follows from the variation of the term $\propto SR$ in the action Eq.~(\ref{eq:action-local}). The fact that $G_{\rm eff}$ depends only on time tells us that gravity is modified with equal strength everywhere, regardless of whether or not one is close to massive bodies or in high-density regions. This may bring into question the ability of this model to pass the stringent Solar System tests of gravity \cite{Will:2014kxa, Babichev:2011iz, 2012PhRvD..85b4023K}. We come back to this discussion in Sec.~\ref{sec:results-ss}. We note also that from Eq.~(\ref{eq:sph2}), it follows that $\Phi = \Psi$ in the quasi-static and weak-field limits.

\subsection{Model parameters}

\begin{figure}
	\centering
	\includegraphics[scale=0.43]{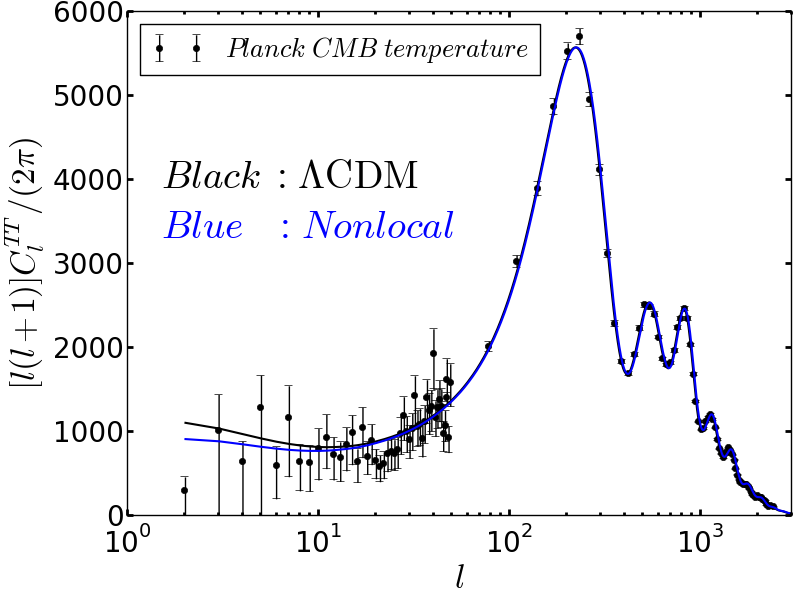}
	\caption{CMB temperature power spectrum of the $\lcdm$ (black) and $\nloc$ (blue) models for the cosmological parameters of Table \ref{table:params}. The data points with errorbars show the power spectrum measured by the Planck  satellite \cite{Ade:2013kta, Ade:2013zuv}.} 
\label{fig:cmb}\end{figure}

\begin{table}
\caption{Cosmological parameter values adopted in this paper. $\Omega_{r0}$, $\Omega_{b0}$, $\Omega_{c0}$, $\Omega_{de0}$, $h$, $n_s$, and $\tau$ are, respectively, the present day fractional energy density of radiation ($r$), baryons ($b$), cold dark matter ($c$) and dark energy ($de$), the dimensionless present day Hubble expansion rate, the primordial scalar spectral index and the optical depth to reionization. The scalar amplitude at recombination $A_s$ refers to a pivot scale $k = 0.05 \rm{Mpc}^{-1}$. These are the $\lcdm$ parameters that best-fit the CMB temperature and lensing data from the Planck satellite \cite{Ade:2013kta, Ade:2013zuv, Ade:2013tyw}, and the BAO data from the 6df \cite{Beutler:2011hx}, SDDS DR7 \cite{Padmanabhan:2012hf} and BOSS DR9 \cite{Anderson:2012sa} galaxy redshift surveys. The parameters were determined by following the strategy outlined in Ref.~\cite{Barreira:2014jha}, although in the latter neutrino masses are also varied in the constraints. For the purpose of this paper we can assume neutrinos to be effectively massless. The $\nloc$ model parameter $m$ is derived by the condition to make the Universe spatially flat, i.e.~, $1 = \Omega_{r0} + \Omega_{b0} + \Omega_{c0} + \Omega_{de0}(m)$.}
\begin{tabular}{@{}lccccccccccc}
\hline\hline
\\
Parameter  & \ \ Planck (temperature+lensing) + BAO& \ \ 
\\
\hline
\\
$\Omega_{r0}{h}^2$                                				&\ \ $4.28\times10^{-5}$ & \ \ 
\\
$\Omega_{b0}{h}^2$                                				&\ \ $0.02219$ & \ \ 
\\
$\Omega_{c0}{h}^2$                                 				&\ \   $0.1177$ & \ \ 
\\
${h}$                                                                			&\ \   $0.6875$ & \ \ 
\\
$n_s$                                                                 			           &\ \    $0.968$ & \ \ 
\\
$\tau$                                                                     			&\ \   $0.0965$ & \ \
\\
$\rm{log}_{10}\left[ 10^{10}A_s \right]$    			&\ \  $3.097$ & \ \ 
\\
\\
\hline
\\
$\Omega_{de0}$                                                                        &\ \   $0.704$     & \ \
\\
$m$                                                                                   &\ \    $0.288$ & \ \
\\
\hline
\hline
\end{tabular}
\label{table:params}
\end{table}

The results presented in this paper are for the cosmological parameter values listed in Table \ref{table:params}. These are the best-fitting $\lcdm$ parameters to a dataset that comprises the CMB data from the Planck satellite (both temperature and lensing) \cite{Ade:2013kta, Ade:2013zuv, Ade:2013tyw}, and the BAO data from the 6df \cite{Beutler:2011hx}, SDDS DR7 \cite{Padmanabhan:2012hf} and BOSS DR9 \cite{Anderson:2012sa} galaxy redshift surveys. The parameters were found by following the steps outlined in Ref.~\cite{Barreira:2014jha}, although in the latter, neutrino masses are also varied in the constraints. In this paper, however, we treat neutrinos as massless for simplicity.

The CMB temperature power spectra of the $\lcdm$ and $\nloc$ models for the parameters listed in Table \ref{table:params} are shown in Fig.~\ref{fig:cmb}. The $\nloc$ model predictions were obtained with a suitably modified version of the {\tt CAMB} code \cite{camb_notes}. The derivation of the perturbed equations that enter the calculations in {\tt CAMB} follows the steps of Ref.~\cite{Barreira:2012kk}, to which we refer the interested reader for details. The results in Fig.~\ref{fig:cmb} shows that the $\nloc$ model is able to fit the CMB data with a goodness-of-fit that is similar to that of $\lcdm$. In fact, the $\nloc$ model is in slightly better agreement with the data at low-$l$, which is mostly determined by the Integrated Sachs-Wolfe (ISW) effect. However, the larger errorbars on these scales due to cosmic variance do not allow stringent constraints to be derived.

In Sec.~\ref{sec:results} we shall compare the results of the $\nloc$ model with those of standard $\lcdm$. In this paper, we are mostly interested in the phenomenology driven by the modifications to gravity in the $\nloc$ model. This is why we shall use the same cosmological parameters for both models. A formal exploration of the constraints on the parameter space in the $\nloc$ model is beyond the scope of the present paper (see Ref.~\cite{swissgroup}).

\section{Halo Model of the nonlinear matter power spectrum}\label{sec:halo-model-formulae}

In this section, we describe the halo model of the nonlinear matter power spectrum, as well as all of its ingredients. In particular, we define the halo mass function, linear halo bias and halo density profiles. We also present the equations that govern the linear growth and spherical collapse of structures.

\subsection{Halo model}\label{sec:halomodel}

In the halo model approach, one assumes that all matter in the Universe lies within gravitationally bound structures (see Ref.~\cite{Cooray:2002dia} for a review). As a result, the two-point correlation function of the matter field can be decomposed into the contributions from the correlations between elements that lie in the same halo (the 1-halo term) and in different haloes (the 2-halo term). The power spectrum can also be decomposed in a similar way, and one can write

\bq\label{eq:halo-model}
P_k = P_k^{\rm 1h} + P_k^{\rm 2h},
\eq
where

\bq\label{eq:halo-model-terms}
P_k^{\rm 1h} &=& \int {\rm d}M \frac{M}{\bar{\rho}_{m0}^2}\frac{{\rm d}n(M)}{{\rm dln}M} |u(k, M)|^2, \nonumber \\
P_k^{\rm 2h} &=& I(k)^2P_{k,{\rm lin}} ,
\eq
are, respectively, the 1- and 2-halo terms, with 

\bq\label{eq:I-function}
I(k) = \int {\rm d}M \frac{1}{\bar{\rho}_{m0}}\frac{{\rm d}n(M)}{{\rm dln}M} b_{\rm lin}(M) |u(k, M)|.
\eq
In Eqs.(\ref{eq:halo-model})-(\ref{eq:I-function}), $\bar{\rho}_{m0}$ is the present-day background (total) matter density; $P_{k, \rm lin}$ is the matter power spectrum obtained using linear theory; $k$ is the comoving wavenumber; ${{\rm d}n(M)}/{{\rm dln}M}$ is the {\it mass function}, which describes the comoving number density of haloes per differential logarithmic interval of mass; $u(k, M)$ is the Fourier transform of the density profile of the haloes truncated at their size and normalized such that $u(k \rightarrow 0, M) \rightarrow 1$; $b_{\rm lin}(M)$ is the linear halo bias parameter. We model all these quantities in the remainder of this section, in which we follow the notation of Refs.~\cite{Barreira:2013xea, 2014JCAP...04..029B}.

\subsection{Halo mass function}\label{sec:halomodel-mf}

We define the halo mass function as

\bq\label{eq:mass-function}
&&\frac{{\rm d}n(M)}{{\rm d}{\rm ln}M}{\rm d}{\rm ln}M = \frac{\bar{\rho}_{m0}}{M} f(S){\rm d}S,
\eq
where $S$ is the variance of the linear density field filtered on a comoving length scale $R$,

\bq\label{eq:variance}
S(R) \equiv \sigma^2(R) = \frac{1}{2\pi^2}\int k^2 P_{k, {\rm lin}}\tilde{W}^2\left(k, R\right){\rm d}k.
\eq
Here, $\tilde{W}\left(k, R\right) = 3\left({\rm sin}(kR) - kR{\rm cos}(kR)\right)/\left(kR\right)^3$ is the Fourier transform of the filter, which we take as a top-hat in real space. The total mass enclosed by the filter is given by

\bq\label{eq:mass-overdensity}
M = 4\pi \bar{\rho}_{m0}R^3/3.
\eq
In Eq.~(\ref{eq:mass-function}), $f(S){\rm d}S$ describes the fraction of the total mass that resides in haloes whose variances lie within $\left[S, S + {\rm d}S\right]$ (or equivalently, whose masses lie within $\left[M-{\rm d}M, M \right]$) \footnote{Note that the quantities $S$, $R$ and $M$ can be related to one another via Eqs.~(\ref{eq:variance}) and (\ref{eq:mass-overdensity}). In this paper, we use these three quantities interchangeably when referring to the scale of the haloes.}. Here, we use the Sheth-Tormen expression \cite{Sheth:1999mn, Sheth:1999su, Sheth:2001dp},

\bq\label{eq:first-crossing-ST}
f(S) = A \sqrt{\frac{q}{2\pi}}\frac{\delta_c}{S^{3/2}}\left[1 + \left(\frac{q\delta_c^2}{S}\right)^{-p}\right]{\rm exp}\left[-q\frac{\delta_c^2}{2S}\right], \nonumber \\
\eq
where $A$ is a normalization constant fixed by the condition $\int f(S){\rm d}S = 1$. $\delta_c \equiv \delta_c(z)$ is the critical initial overdensity for a spherical top-hat to collapse at redshift $z$, extrapolated to $z = 0$ with the $\Lambda$CDM linear growth factor. This extrapolation is done purely to ensure that the resulting values of $\delta_c$ can be more readily compared to values in $\lcdm$. Note that for consistency, $P_{k, {\rm lin}}$ in Eq.~(\ref{eq:variance}) is also the initial power spectrum of the specific model, evolved to $z = 0$ with the $\Lambda$CDM linear growth factor.

The Press-Schechter mass function \cite{1974ApJ...187..425P} is obtained by taking $(q,p) = (1,0)$ in Eq.~(\ref{eq:first-crossing-ST}). This choice is motivated by the spherical collapse model. However, Refs.~\cite{Sheth:1999mn, Sheth:1999su, Sheth:2001dp} showed that the choice of parameters $(q,p) = (0.75,0.3)$ (motivated by the ellipsoidal, instead of spherical collapse) leads to a better fit to the mass function measured from N-body simulations of $\Lambda$CDM models (see also Ref.~\cite{2010ApJ...717..515M}). For alternative models, such as those with modified gravity, it is not necessarily true that the standard ST parameters ($(q,p) = (0.75, 0.3)$) also provide a good description of the simulation results. For example, in Ref.~\cite{2014JCAP...04..029B}, we demonstrated that the ST mass function can only provide a good fit to the simulation results of Galileon gravity models \cite{PhysRevD.79.064036, DeFelice:2010pv, Deffayet:2009mn, Barreira:2012kk} after a recalibration of the $(q,p)$ parameters. In Sec.~\ref{sec:results-mf}, we shall investigate the need for a similar recalibration in the $\nloc$ model.

\subsection{Linear halo bias}\label{sec:halomodel-hb}

The linear halo bias parameter $b(M)$ \cite{Mo:1995cs} relates the clustering amplitude of haloes of mass $M$ to that of the total matter field on large length scales ($k \ll 1 h/{\rm Mpc}$),

\bq\label{eq:bias-def}
\delta_{\rm halo}(M) = b(M)\delta_{\rm matter},
\eq
where $\delta_{\rm halo}$ and $\delta_{\rm matter}$ are the density contrast of the distribution of haloes of mass $M$ and of the total matter field, respectively. On smaller length scales, where the matter overdensities become larger $\delta_{\rm matter} \gtrsim 1$, Eq.~(\ref{eq:bias-def}) requires higher order corrections (see e.g.~\cite{Fry:1992vr}).

Following the same derivation steps as in Ref.~\cite{Barreira:2013xea}, it is straightforward to show that the ST linear halo bias parameter can be written as

\bq\label{eq:st-halo-bias}
b(M) = 1 + {g(z)}\left(\frac{q{\delta_c^2}/{S} - 1}{\delta_c} + \frac{2p/\delta_c}{1 + \left(q\delta_c^2/S\right)^p}\right),
\eq
with $g(z) = D^{\Lambda \rm CDM}(z = 0) / D^{\rm Model}(z)$, where $D(z)$ is the linear growth factor of a specific model defined as $\delta_{\rm matter}(z) = D(z)\delta_{\rm matter}(z_i)/D(z_i)$.

\subsection{Halo density profiles}\label{sec:halomodel-prof}

We adopt the NFW formula  \cite{Navarro:1996gj} to describe the radial density profile of dark matter haloes

\bq\label{eq:nfw}
\rho_{\rm NFW}(r) = \frac{\rho_s}{{r}/{r_s}\left[1 + {r}/{r_s}\right]^2},
\eq
where $\rho_s$ and $r_s$ are often called the {\it characteristic density} and the {\it scale radius} of the halo. The mass of the NFW halo, $M_{\Delta}$, is obtained by integrating Eq.~(\ref{eq:nfw}) up to some radius $R_{\Delta}$ (the meaning of the subscript $\Delta$ will become clear later)

\bq\label{eq:mass-nfw}
M_{\Delta} &=& \int_0^{R_{\Delta}}{\rm d}r4\pi r^2 \rho_{\rm NFW}(r) \\ \nonumber
&=& 4\pi \rho_s\frac{R_{\Delta}^3}{c_{\Delta}^3}\left[{\rm ln}\left(1+c_{\Delta} \right) - \frac{c_{\Delta}}{1+c_{\Delta}}\right],
\eq
where $c_{\Delta} = {R_{\Delta}}/{r_s}$ is the concentration parameter.

In our simulations, we define the halo mass as

\bq\label{eq:mass-sim}
M_{\Delta} = \frac{4\pi}{3}\Delta\bar{\rho}_{c0}R_{\Delta}^3,
\eq
i.e., $M_{\Delta}$ is the mass that lies inside a comoving radius $R_{\Delta}$, within which the mean density is $\Delta$ times the critical density of the Universe at the present day, $\bar{\rho}_{c0}$. In this paper, we take $\Delta= 200$. By combining the two mass definitions of Eqs.~(\ref{eq:mass-nfw}) and (\ref{eq:mass-sim}), one gets $\rho_s$ as a function of $c_{\Delta}$:

\bq\label{eq:rhos-nfw}
\rho_s = \frac{1}{3}\Delta\bar{\rho}_{c0} c_{\Delta}^3\left[{\rm ln}\left(1+c_{\Delta} \right) - \frac{c_{\Delta}}{1+c_{\Delta}}\right]^{-1}.
\eq
The NFW profile then becomes fully specified by the values of $r_s$, which are determined by direct fitting to the density profiles of the haloes from the simulations. Equivalently, and as is common practice in the literature, one can specify the concentration-mass relation $c_{\Delta}(M_{\Delta})$, instead of $r_s(M_{\Delta})$. In the context of $\lcdm$ cosmologies, the (mean) concentration-mass relation is well fitted by a power law function over a certain mass range \cite{Bullock:1999he, Neto:2007vq, Maccio':2008xb, Prada:2011jf}. The same is true, for instance, for Galileon gravity models \cite{2014JCAP...04..029B}, although with very different fitting parameters. A proper assessment of the performance of the halo model prescription therefore requires us to fit the concentration-mass relation of the $\nloc$ simulations as well. This is done in Sec.~\ref{sec:results-c}. Given the relation $c_{\Delta}(M_{\Delta})$, then the NFW density profile becomes completely specified by the halo mass $M_{\Delta}$.

Finally, since it is the Fourier transform of the profiles, $u(k, M)$, and not the profiles themselves, that enter Eqs.~(\ref{eq:halo-model-terms}) and (\ref{eq:I-function}), we simply mention that it is possible to show that

\bq\label{eq:nfw-fourier}
&&u_{\rm NFW}(k, M) = \int_0^{R_\Delta} {\rm d}r4\pi r^2\frac{{\rm sin}kr}{kr}\frac{\rho_{\rm NFW}(r)}{M_{\Delta}} \nonumber \\
&=& {4\pi \rho_s r_s^3}\left\{\frac{{\rm sin}\left(kr_s\right)}{M}\left[{\rm Si}\left(\left[1+c_{\Delta}\right]kr_s\right) - {\rm Si}\left( kr_s\right)\right]\right. \nonumber \\
&&\ \ \ \ \ \ \ \ \ \ \ + \left.\frac{{\rm cos}\left(kr_s\right)}{M}\left[{\rm Ci}\left(\left[1+c_{\Delta}\right]kr_s\right) - {\rm Ci}\left( kr_s\right)\right]\right. \nonumber \\
&&\ \ \ \ \ \ \ \ \ \ \  - \left.  \frac{{\rm sin}\left(c_\Delta kr_s\right)}{M\left(1+c_\Delta\right)kr_s}\right\},
\eq
where ${\rm Si}(x) = \int_0^x {\rm d}t{\rm sin}(t)/t$ and ${\rm Ci}(x) = -\int_x^\infty {\rm d}t{\rm cos}(t)/t$. Note that $u(k \rightarrow 0,M) \rightarrow 1$, as required by its normalization.

\subsection{Linear growth factor and spherical collapse dynamics}\label{sec:halomodel-dyn}

In order to use the ST formulae for the mass function and linear halo bias one still needs to specify and solve the equations that determine the threshold density $\delta_c$ and the evolution of the linear overdensities. For scales well within the horizon, the linear (small) density contrast $\delta_{\rm lin}$ is governed by

\bq\label{eq:linear-growthfactor0}
\ddot{\delta}_{\rm lin} + 2H\dot{\delta}_{\rm lin} - 4\pi {G_{{\rm eff}}(a)}\bar{\rho}_m\delta_{\rm lin} = 0,
\eq
or equivalently, by changing the time variable to $N = {\rm ln} a$, by

\bq\label{eq:linear-growthfactor}
D'' + \left(\frac{E'}{E} + 2\right)D' - \frac{3}{2}\frac{G_{{\rm eff}}(a)}{G}\frac{\Omega_{m0}e^{-3N}}{E^2} &=& 0,
\eq
where we have used that $\delta_{\rm lin}(a) = D(a)\delta_{\rm lin}(a_i)/D(a_i)$ and a prime denotes a derivative w.r.t. $N$. The initial conditions are set up at $z_i = 300$ using the known matter dominated solution $D(a_i) = D'(a_i) = a_i$. The $\nloc$ model changes the way structure grows on large scales compared to $\lcdm$ via its modifications to $E(a) \equiv H(a)/H_0$ and $G_{\rm eff}/G$. 

We have defined $\delta_c(z)$ as the linearly extrapolated value (using the $\Lambda$CDM linear growth factor) of the initial overdensity of a spherical region for it to collapse at a given redshift, $z$. To determine $\delta_c$, we consider the evolution equation of the physical radius $\zeta = a(t)r$ of the spherical halo at time $t$, which satisfies the Euler equation

\bq\label{eq:sph_col}
\frac{\ddot{\zeta}}{\zeta} - \left(\dot{H} + H^2\right) &=& -\frac{\Phi,_{\zeta}}{\zeta} = -\frac{G_{{\rm eff}}(a)}{G}\frac{{H_0^2}\Omega_{m0}\delta a^{-3}}{2},
\eq
where the last equality follows from integrating Eq.~(\ref{eq:modpoisson}) over $\int r^2{\rm d}r$. Changing the time variable to $N$ and defining $y(t) = \zeta(t)/\left(aR\right)$, Eq.~(\ref{eq:sph_col}) becomes

\bq\label{eq:sph_col1}
y'' &+& \left(\frac{E'}{E} + 2\right)y' \nonumber \\
&+& \frac{G_{{\rm eff}}(a)}{G}\frac{\Omega_{m0}e^{-3N}}{2E^2}\left(y^{-3} - 1\right)y = 0,
\eq
where we have used $\delta = y^{-3} - 1$, which follows from mass conservation \footnote{Explicitly, one has $\bar{\rho}_ma^3R^3 = \left(1+\delta\right)\bar{\rho}_m\zeta^3 \Rightarrow \delta = \left(aR/\zeta\right)^3 - 1 = y^{-3} - 1$.}. The initial conditions are set up as $y(a_i) = 1 - \delta_{\rm lin, i}/3$ and $y'(a_i) = \delta_{\rm lin,i}/3$ (here, $\delta_{\rm lin,i}$ is the linear density contrast at the initial time). The value of $\delta_c$ is then determined by finding the value of the initial density $\delta_{\rm lin, i}$ that leads to collapse ($y = 0$, $\delta \rightarrow \infty$) at redshift $z$, evolving this afterwards until today using the $\Lambda$CDM linear growth factor.

As we have noted above, in the $\nloc$ model, the modifications to gravity are time dependent only. In other words, the value of $G_{\rm eff}$ is the same on large and on small scales. This is different, for instance, from the case of Galileon gravity. In the latter, the nonlinearities of the Vainshtein screening mechanism suppress the effective gravitational strength felt by a test particle that lies within a certain radius (known as Vainshtein radius) from a matter source. In Eq.~(\ref{eq:sph_col1}), this would be simply taken into account by replacing $G_{\rm eff}(a)$ with $G_{\rm eff}(a, \delta = y^{-3} - 1)$ \cite{Barreira:2013xea}. The picture becomes more complicated in the case of modified gravity models which employ chameleon-type screening mechanisms \cite{Khoury:2003aq, Khoury:2003rn}. In these models, $G_{\rm eff}$ also depends on the size (or mass) of the halo and on the gravitational potentials in the environment where the halo forms. This requires a generalization of the spherical collapse formalism for these models, which has been developed by Ref.~\cite{Li:2011qda}.

\section{Results}\label{sec:results}

\subsection{N-body simulations summary}\label{sec:results-simsum}

\begin{table}
\caption{Summary of the three models we simulate in this paper. All models share the cosmological parameters of Table \ref{table:params}. The $\qcdm$ model has the same expansion history as the $\nloc$ model, but with GR as the theory of gravity (cf.~Sec.~\ref{sec:results-simsum}).}
\begin{tabular}{@{}lccccccccccc}
\hline\hline
\\
Model  & \ \ $H(a)$ & $G_{\rm eff}/G$ &\ \ 
\\
\hline
\\
"Full" $\nloc$      &\ \  $H(a)_{\nloc}$ &  \ \ Eq.~(\ref{eq:geff}) &\ \ 
\\
$\qcdm$               &\ \ $H(a)_{\nloc}$ &  $1$ &\ \ 
\\
$\lcdm$                &\ \ $H(a)_{\lcdm}$ &  $1$&\ \ 
\\
\hline
\hline
\end{tabular}
\label{table:models}
\end{table}

Our simulations were performed with a modified version of the publicly available {\tt RAMSES} N-body code \cite{Teyssier:2001cp}. {\tt RAMSES} is an Adaptive Mesh Refinement (AMR) code, which solves the Poisson equation on a grid that refines itself when the effective number of particles within a given grid cell exceeds a user-specified threshold, $N_{\rm th}$. Our modifications to the code consist of (i) changing the routines that compute the background expansion rate to interpolate the $\nloc$ model expansion rate from a pre-computed table generated elsewhere; (ii) re-scaling the total force felt by the particles in the simulation by $G_{\rm eff}(a)/G$, whose values are also interpolated from a table generated beforehand. 

In the following sections we show the N-body simulation results obtained for three models. We simulate the "full" $\nloc$ model of action Eq.~(\ref{eq:action}), whose expansion history and $G_{\rm eff}/G$ are given by Eqs.~(\ref{eq:friedmann1}) and (\ref{eq:geff}), respectively. We also simulate a standard $\lcdm$ model and a model with the same expansion history as the $\nloc$ model, but with $G_{\rm eff}/G = 1$. We call the latter model $\qcdm$, and comparing its results to $\lcdm$ allows us to pinpoint the impact of the modified $H(a)$ alone on the growth of structure. The specific impact of the modified $G_{\rm eff}$ can then be measured by comparing the results from the "full" $\nloc$ model simulations with those from $\qcdm$. Table \ref{table:models} summarizes the models we consider in this paper. 

We simulate all models on a cubic box of size $L = 200\ {\rm Mpc}/h$ with $N_{p} = 512^3$ dark matter particles. We take $N_{\rm th} = 8$ as the grid refinement criterion. The initial conditions are set up at $z = 49$, using the $\lcdm$ linear matter power spectrum with the parameters of Table \ref{table:params}. For each model, we simulate five realizations of the initial conditions (generated using different random seeds), which we use to construct errorbars for the simulation results by determining the variance across the realizations.

Finally, we simply note that the modifications to {\tt RAMSES} needed to simulate the $\nloc$ model are trivial compared to those that are necessary to simulate models such as $f(R)$ \cite{Li:2011vk, Puchwein:2013lza, Llinares:2013jza, Oyaizu:2008sr, 2011PhRvD..83d4007Z, Hellwing2013b}, Symmetron \cite{2012ApJ...748...61D, Llinares:2013jua} or Galileons \cite{2009PhRvD..80j4005C, Schmidt:2009sg, Li:2013nua, Wyman:2013jaa, Barreira:2013eea, Li:2013tda}. In these, because of the screening mechanisms (which introduce density and scale dependencies of the total force), additional solvers are needed for the (nonlinear) equations of the extra scalar degrees of freedom. In the {\tt ECOSMOG} code \cite{Li:2011vk} (also based on {\tt RAMSES}), these equations are solved via Gauss-Seidel relaxations on the AMR grid, which makes the simulations significantly more time consuming. We note also that recently, Ref.~\cite{Winther:2014cia} has proposed a new and faster scheme to simulate screened modified gravity in the mildly non-linear regime. This scheme uses the linear theory result, but combines it with a screening factor computed analytically assuming spherical symmetry, which helps speed up the calculations without sacrificing the accuracy on mildly nonlinear scales too much.

\subsection{Linear growth and $\delta_c$ curves}\label{sec:results-lingf}

Before discussing the results from the simulations, it is instructive to look at the model predictions for the linear growth rate of structure and for the time dependence of the critical density $\delta_c(z)$.

\begin{figure}
	\centering
	\includegraphics[scale=0.35]{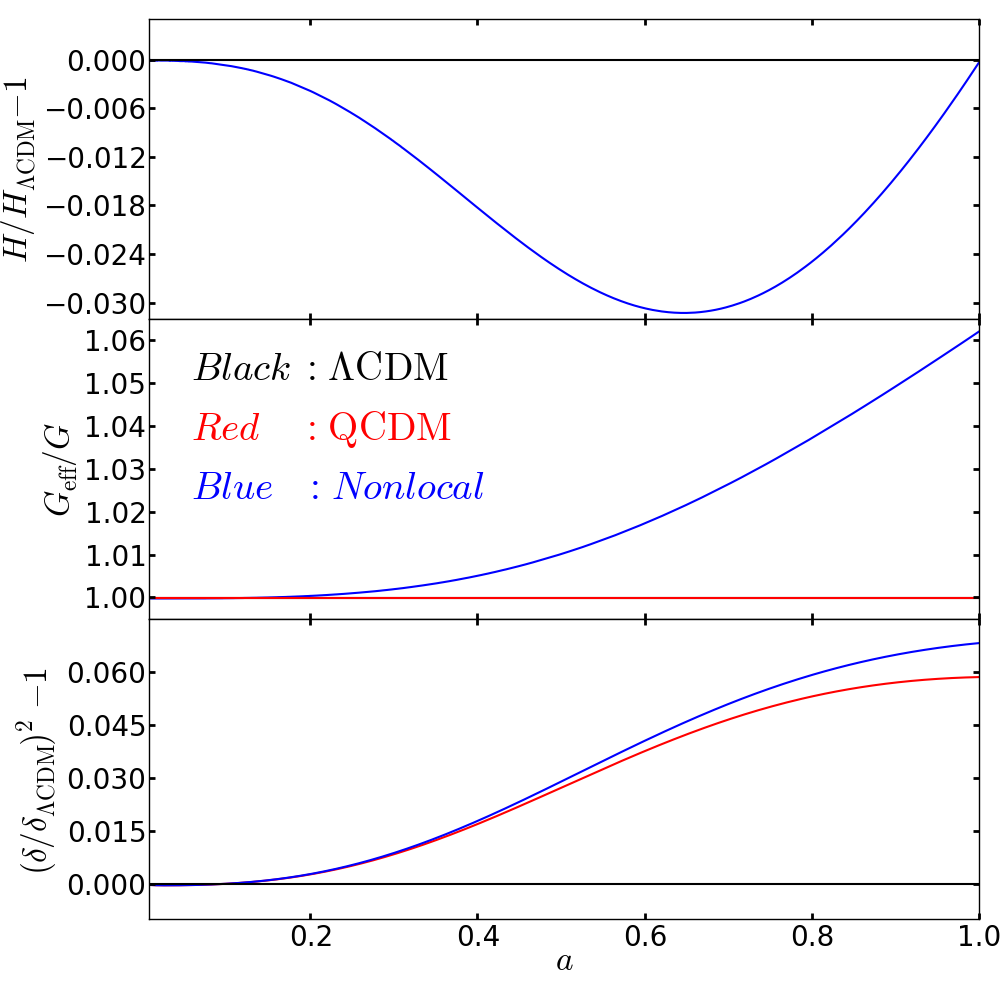}
	\caption{The upper panel shows the evolution of the expansion rate, plotted as the fractional difference w.r.t. the $\lcdm$ (black) result, $H(a)/H_{\lcdm} - 1$ as a function of the expansion factor, $a$. $H(a)$ is the same for the $\nloc$ (blue) and $\qcdm$ (red) models. The middle panel shows the evolution of the effective gravitational strength, $G_{\rm eff}/G$. This is unity in the $\lcdm$ and $\qcdm$ models at all times. The lower panel shows the evolution of the squared linear density contrast, $\delta^2$, plotted as the fractional difference w.r.t. the $\lcdm$ prediction.} 
\label{fig:lingf}\end{figure}

\begin{figure}
	\centering
	\includegraphics[scale=0.38]{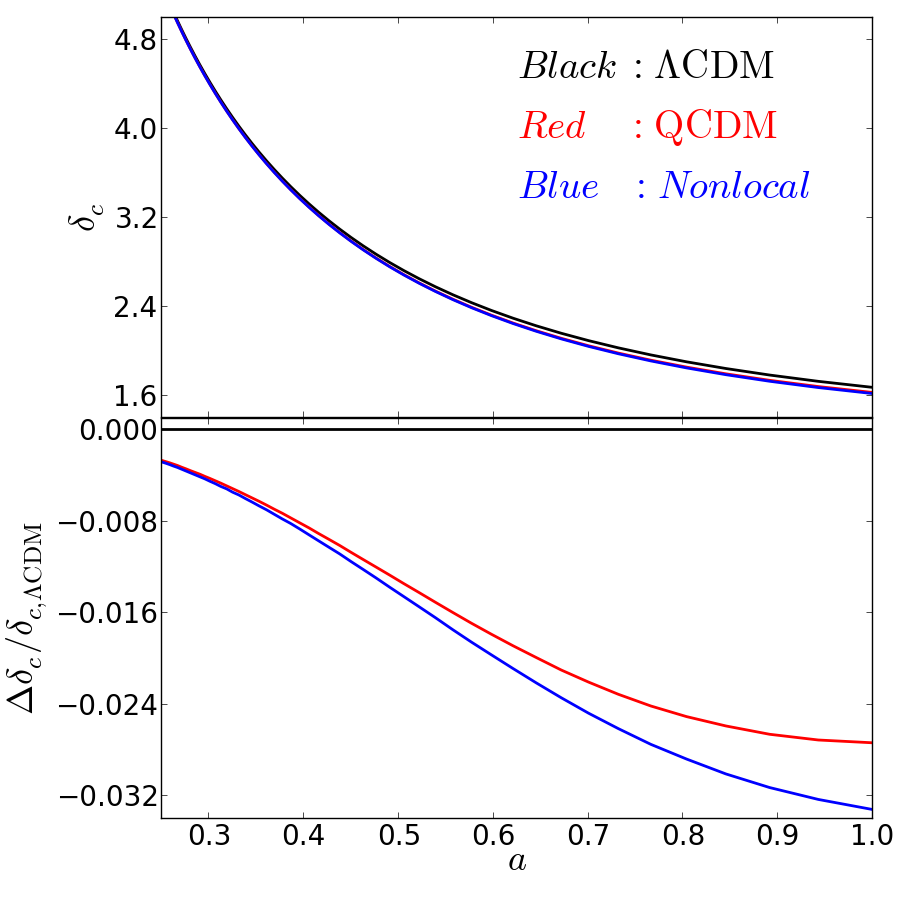}
	\caption{The upper panel shows the time evolution of the critical initial density for a spherical top-hat halo to collapse at scale factor $a$ (linearly extrapolated to the present-day using the $\lcdm$ linear growth factor), for the $\lcdm$ (black), $\qcdm$ (red) and $\nloc$ (blue) models. The lower panel shows the fractional difference w.r.t. $\lcdm$.}
\label{fig:dc}\end{figure}

\begin{table}
\caption{Values of the critical initial overdensity for the collapse of a spherical top-hat to occur at $a = 0.6$, $a = 0.8$, $a = 1.0$, extrapolated to $a = 1.0$ with the $\Lambda$CDM linear growth factor. This extrapolation is done to allow the resulting values of $\delta_c$ to be more easily compared to values in $\lcdm$.}
\begin{tabular}{@{}lccccccccccc}
\hline\hline
\\
Model  & \ \ $a = 0.6$ & $a = 0.8$ & $a = 1.0$ &\ \ 
\\
               & \ \ $\delta_c$ & $\delta_c$ & $\delta_c$ &\ \ 
\\
\hline
\\
$\nloc$                       &\ \  $2.316$ &  \ \ $1.859$ & $1.622$&\ \ 
\\
$\qcdm$                    &\ \  $2.320$ &  \ \ $1.866$ & $1.632$&\ \ 
\\
$\lcdm$                     &\ \  $2.362$ &  \ \ $1.913$ & $1.678$&\ \ 
\\
\\
\hline
\hline
\end{tabular}
\label{table:dc}
\end{table}

From top to bottom, Fig.~\ref{fig:lingf} shows the time evolution of the fractional difference of the expansion rate relative to $\lcdm$, $H/H_{\lcdm} - 1$, the effective gravitational strength $G_{\rm eff}/G$ and the fractional difference of the squared linear density contrast relative to $\lcdm$, $\left(\delta/\delta_{\lcdm}\right)^2 - 1$. The expansion rate in the $\nloc$ model is lower than in $\lcdm$ for $a \gtrsim 0.1$. This reduces the amount of Hubble friction and therefore boosts the linear growth rate. The gravitational strength in the $\nloc$ model starts growing after $a \gtrsim 0.2$, being approximately $6\%$ larger than in GR at the present day. This also boosts the linear growth of structure, but has a smaller impact compared to the effect of the lower expansion rate. This is seen by noting that the differences between $\qcdm$ and $\lcdm$ in the bottom panel are larger than the differences between $\qcdm$ and the $\nloc$ model.

Figure \ref{fig:dc} shows the time dependence of $\delta_c$. In the top panel, all models exhibit the standard result that $\delta_c$ decreases with time, i.e., the initial overdensity of the spherical top-hat should be smaller, if the collapse is to occur at later times. Compared to $\lcdm$, at late times ($a \gtrsim 0.3$), the $\qcdm$ and $\nloc$ models predict lower values for $\delta_c$. This is as expected since structure formation is boosted at late times in these models, and as a result, this needs to be compensated by smaller values of the initial overdensities for the collapse to occur at the same epoch as in $\lcdm$. Just like in the case of the linear growth rate, the differences w.r.t. the $\lcdm$ results are mainly affected by the lower expansion rate, and not by the larger values of $G_{\rm eff}/G$. At earlier times ($a \lesssim 0.3$), all models have essentially the same expansion rate and gravitational strength, and as a result, the values of $\delta_c$ are roughly the same. Table \ref{table:dc} shows the values of $\delta_c$ at $a = 0.60$, $a = 0.80$ and $a = 1.00$, for the three models.

\subsection{Interpretation of the constraints from Solar System tests of gravity}\label{sec:results-ss}

The absence of a screening mechanism in the $\nloc$ model may raise concerns about the ability of the model to satisfy Solar System constraints \cite{Will:2014kxa}. For instance, for the parameters of Table \ref{table:params}, the $\nloc$ model predicts that the rate of change of the gravitational strength today, $\dot{G}_{\rm eff}/G$, is

\bq\label{eq:dtg}
\frac{\dot{G}_{\rm eff}}{G} = H_0\frac{{\rm d}}{{\rm d}N}\left(\frac{G_{\rm eff}}{G}\right) \approx 92 \times 10^{-13}\ \rm{yrs}^{-1},
\eq
which is at odds with the observational contraint $\dot{G}_{\rm eff}/G = \left(4 \pm 9 \right)\times 10^{-13}\ \rm{yr}^{-1}$, obtained from Lunar Laser Ranging experiments \cite{Williams:2004qba}.  Hence, it seems that this type of local constraints can play a crucial role in determining the observational viability of the $\nloc$ model, potentially ruling it out (see e.g.~Refs.~\cite{Babichev:2011iz, 2012PhRvD..85b4023K} for a similar conclusion, but in the context of other models).

It is interesting to contrast this result with that of the nonlocal model of Ref.~\cite{Deser:2007jk}, which we call here the $\dw$ model (for brevity), where $f(X)$ is a free function that appears in the action and $X = \Box^{-1}R$. The equations of motion of this model can be schematically written as

\bq\label{eq:fedw}
G_{\mu\nu}\left[1 + \chi(X)\right] + \Delta G_{\mu\nu} = T_{\mu\nu},
\eq
where $\Delta G_{\mu\nu}$ encapsulates all the extra terms that are not proportional to $G_{\mu\nu}$ and the factor $\chi(X)$ is given by

\bq\label{eq:chi}
\chi = f(X) + \Box^{-1}\left[R\frac{{\rm d}f}{{\rm d}X}(X)\right].
\eq
For the purpose of our discussion, it is sufficient to look only at the effect of $\chi$ in Eq.~(\ref{eq:fedw}). This rescales the gravitational strength as

\bq
\frac{G_{\rm eff}}{G} = \left\{1 + \chi\right\}^{-1},
\eq
which is similar to the effect of $S$ in the $\nloc$ model. There is, however, one very important difference associated with the fact that in the case of the $\dw$ model, one has the freedom to choose the functional form of the terms that rescale $G_{\rm eff}$. To be explicit, we write the argument of $f$ as

\bq\label{eq:X}
 X = \Box^{-1}R = \Box^{-1}\bar{R} + \Box^{-1}\delta R,
\eq
where $\bar{R}$ and $\delta R$ are, respectively, the background and spatially perturbed part of $R$. As explained in Ref.~\cite{Woodard:2014iga}, the relative size of $\bar{R}$ and $\delta R$ is different in different regimes. At the background level, $\Box^{-1}\delta R = 0$ and so the operator $\Box^{-1}$ acts only on $\bar{R}$. On the other hand, within gravitationally bound objects we have $\Box^{-1}\delta R > \Box^{-1}\bar{R}$. Now recall that the covariant $\Box$ operator acts with different signs on purely time- and space-dependent quantities \footnote{For instance, in flat four-dimensional Minkowski space we have $\Box = +\frac{\partial^2}{\partial t^2} - \frac{\partial^2}{\partial x^2} - \frac{\partial^2}{\partial y^2} - \frac{\partial^2}{\partial z^2}$.}. As a result, the sign of $X$ on the background differs from that within bound systems, such as galaxies or our Solar System. This can be exploited to tune the function $f$ in such a way that it vanishes when the sign of $X$ is that which corresponds to bound systems. In this way, $\chi = 0$ and one recovers GR completely \footnote{In Eq.~(\ref{eq:fedw}), $\Delta G_{\mu\nu}$ also vanishes if $\chi = 0$.}. When $X$ takes the sign that corresponds to the background, then the function $f$ is tuned to reproduce a desired expansion history, typically $\lcdm$. In the case of the $\nloc$ model, $S$ is fixed to be $S = \Box^{-2}\left(\bar{R} + \delta R\right)$ and one does not have the freedom to set it to zero inside bound objects. Consequently, the time-dependent part of $S$ is always present in Eq.~(\ref{eq:geff}), which could potentially lead to a time-dependent gravitational strength that is at odds with the current constraints.

For completeness, one should be aware of a caveat. In the above reasoning, we have always assumed that the line element of Eq.~(\ref{eq:ds}) is a good description of the geometry of the Solar System. The question here is whether or not the factor $a(t)^2$ should be included in the spatial sector of the metric when describing the Solar System. This is crucial as the presence of $a(t)$ in Eq.~(\ref{eq:ds}) determines if $S$ varies with time or not. If $a(t)$ is considered, then $S$ varies with time and $G_{\rm eff}$ is time-varying as well. In this way, the model fails the Solar System tests. On the other hand, if one does not consider $a(t)$ in the metric, then $G_{\rm eff}$ is forcibly constant, and there are no apparent observational tensions. Such a static analysis was indeed performed by Refs.~\cite{Kehagias:2014sda, Maggiore:2014sia}, where it was shown that the model can cope well with the local constraints.

This boils down to determining the impact of the global expansion of the Universe on local scales. It is not clear to us that if a field is varying on a time-evolving background, then it should not do so in a small perturbation around that background. However, we acknowledge this is an open question to address, and such study is beyond the scope of the present paper. In what follows, we limit ourselves to assuming that Eq.~(\ref{eq:geff}) holds on all scales, but focus only on the cosmological (rather than local) interpretation of the results.

\subsection{Halo mass function}\label{sec:results-mf}

\begin{figure*}
	\centering
	\includegraphics[scale=0.39]{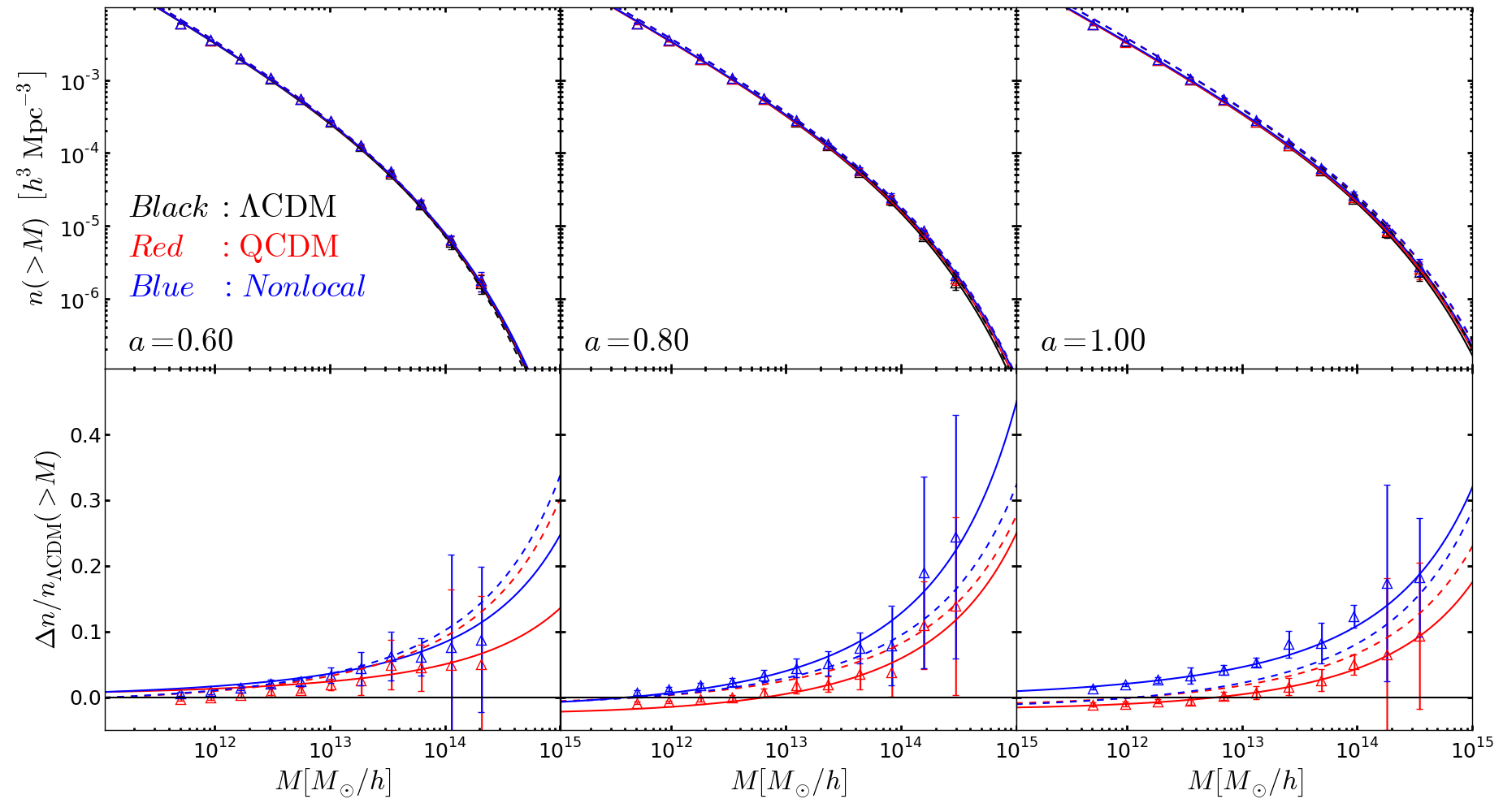}
	\caption{The cumulative mass function of dark matter haloes (upper panels) for the $\lcdm$ (black), $\qcdm$ (red) and $\nloc$ (blue) models, at three epochs $a = 0.6$, $a = 0.8$ $a = 1.0$, as labelled. The lower panels show the difference w.r.t. the $\lcdm$ model results. The symbols show the simulation results, and the errorbars indicate twice the variance across the five realizations of the initial conditions. We have used the phase-space friends-of-friends {\tt Rockstar} code \cite{Behroozi:2011ju} to build the halo catalogues (without subhalos) used to compute the halo abundances. We only show the results for haloes with mass $M_{200} > 100\times M_p \sim 5\times 10^{11} \msun$, where $M_p = \rho_{m0}L^3/N_p$ is the particle mass in the simulations. The lines correspond to the ST mass function of Eqs.~(\ref{eq:mass-function}) and (\ref{eq:first-crossing-ST}) obtained using the fitted (solid lines) and the standard (dashed lines) $(q,p)$ parameters listed in Table \ref{table:qp}.}
\label{fig:mf}\end{figure*}

\begin{table}
\caption{Best-fitting Sheth-Tormen $(q,p)$ parameters to the simulation results at $a = 0.6$, $a = 0.8$ and $a = 1.0$. The uncertainty in the values of $q$ and $p$ is $\Delta_q = 3.5\times10^{-3}$ and $\Delta_p = 1.5\times 10^{-3}$, respectively. These parameters are those that minimize the quantity $\sum_i |n^{\rm sims}(>M_i)/n^{\rm ST}(>M_i, q, p) - 1|$, in which $n^{\rm sims}$ is the cumulative mass function measured from the simulations and $n^{\rm ST}$ is the analytical result given by the Sheth-Tormen mass function of Eqs.~(\ref{eq:mass-function}) and (\ref{eq:first-crossing-ST}). Here, the index $i$ runs over the number of bins used in the simulation results.}
\begin{tabular}{@{}lccccccccccc}
\hline\hline
\\
Model  & \ \ $a = 0.6$ & $a = 0.8$ & $a = 1.0$&\ \ 
\\
              & \ \ $(q,p)$ & $(q,p)$ & $(q,p)$&\ \ 
\\
\hline
\\
Standard                    &\ \  $(0.750, 0.300)$ &  \ \ $(0.750, 0.300)$ & $(0.750, 0.300)$&\ \ 
\\
\\
$\lcdm$                    &\ \  $(0.713, 0.323)$ &  \ \ $(0.756 , 0.326)$ & $(0.756, 0.341)$&\ \ 
\\
$\qcdm$                    &\ \  $(0.727, 0.321)$ &  \ \ $(0.756, 0.331)$ & $(0.763, 0.344)$&\ \ 
\\
$\nloc$                        &\ \  $(0.720, 0.321)$ &  \ \ $(0.741, 0.326)$ & $(0.756, 0.336)$&\ \ 
\\
\hline
\hline
\end{tabular}
\label{table:qp}
\end{table}

Our results for the cumulative mass function of the $\lcdm$ (black), $\qcdm$ (red) and $\nloc$ (blue) models are shown in Fig.~\ref{fig:mf} at $a = 0.60$, $a = 0.80$ and $a = 1.00$. The symbols show the simulation results obtained with the halo catalogues we built using the {\tt Rockstar} halo finder \cite{Behroozi:2011ju}. The results in the figure correspond to catalogues with subhaloes filtered out. The lines show the ST analytical prediction (Eqs.~(\ref{eq:mass-function}) and (\ref{eq:first-crossing-ST})) computed for the fitted (solid lines) and standard (dashed lines) ST $(q,p)$ parameters of Table \ref{table:qp}. The $(q,p)$ parameters were fitted for all of the epochs shown, using the corresponding values of $\delta_c$ in Table \ref{table:dc}. From the figure, one notes that although performing the fitting helps to improve the accuracy of the analytical formulae, overall the use of the standard values for $(q,p)$ provides a fair estimate of the halo abundances in the $\nloc$ model, and of its relative difference w.r.t. $\lcdm$. This is not the case, for instance, in Galileon gravity models, for which Ref.~\cite{2014JCAP...04..029B} has found that it is necessary to recalibrate substantially the values of $(q,p)$ if the ST mass function is to provide a reasonable estimate of the effects of the modifications to gravity. In the case of the $\nloc$ model, the fact that the standard values of $(q,p) = (0.75, 0.30)$ work reasonably well means that the modifications in the $\nloc$ model, relative to $\lcdm$, are mild enough for its effects on the mass function to be well captured by the differences in $\delta_c(z)$.

At $a = 1.00$, the mass function of the $\qcdm$ model shows an enhancement at the high-mass end ($M \gtrsim 5 \times 10^{12} \msun$), and a suppression at the low-mass end ($M \lesssim 5 \times 10^{12} \msun$), relative to $\lcdm$. This is what one expects in hierarchical models of structure formation if the growth rate of structure is boosted, as smaller mass objects are assembled more efficiently to form larger structures, leaving fewer of them. The effects of the enhanced $G_{\rm eff}/G$ maintain  this qualitative picture, but change it quantitatively. More explicitly, the mass scale below which the mass function drops below that of $\lcdm$ is smaller than the mass range probed by our simulations; and the enhancement of the number density of massive haloes is more pronounced. In particular, compared to $\lcdm$, haloes with masses $M \sim 10^{14} \msun$ are $\sim 5\%$ and $\sim 15\%$ more abundant in the $\qcdm$ and $\nloc$ models, respectively. Figure \ref{fig:mf} also shows that the relative differences w.r.t. $\lcdm$ do not change appreciably with time after $a \sim 0.80$. At earlier times ($a \sim 0.60$), the halo abundances in the $\qcdm$ and $\nloc$ models approach one another, and their relative difference to $\lcdm$ decreases slightly, compared to the result at later times.

\subsection{Halo bias}\label{sec:results-hb}

\begin{figure*}
	\centering
	\includegraphics[scale=0.39]{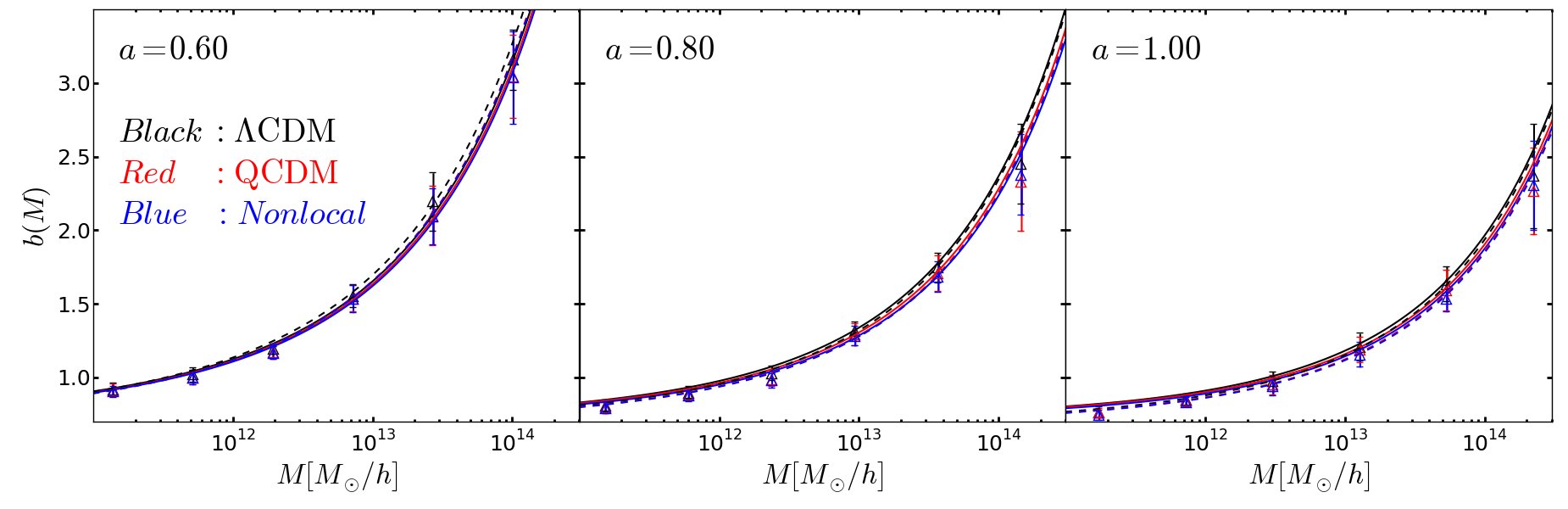}
	\caption{Linear halo bias in the $\lcdm$ (black), $\qcdm$ (red) and $\nloc$ (blue) models, for three epochs $a = 0.6$, $a = 0.8$ and $a = 1.0$, as labelled. The symbols show the asymptotic value of the halo bias on large scales measured from the simulations as $b(M) = P_{\rm hm}(k \rightarrow 0, M)/P(k)$, considering only haloes (and not subhaloes) with mass $M_{200} > 100\times M_p \sim 5\times 10^{11} \msun$, where $M_p = \rho_{m0}L^3/N_p$ is the particle mass. Only the mass bins for which the values of $P_{\rm hm}(k, M)/P(k)$ have reached a constant value on large scales are shown. The errorbars show twice the variance across the five realizations of the initial conditions. The solid and dashed lines show the prediction from the ST formula, Eq.~(\ref{eq:st-halo-bias}), computed, respectively, with the best-fitting and standard $(q,p)$ parameters listed in Table \ref{table:qp}.}
\label{fig:bias}\end{figure*}

The linear halo bias predictions for the $\lcdm$ (black), $\qcdm$ (red) and $\nloc$ (blue) models are shown in Fig.~\ref{fig:bias} at $a = 0.6$, $a = 0.8$ and $a = 1.0$. The symbols show the simulation results, which were obtained by measuring the ratio

\bq\label{eq:hb-sims}
b(k, M) = \frac{P_{\rm hm}(k, M)}{P_k},
\eq
where $P_k$ is the total matter power spectrum and $P_{\rm hm}(k, M)$ is the halo-matter cross spectrum for haloes of mass $M$. These were measured with the aid of a {\it Delaunay Tessellation} field estimator code \cite{cv2011, sv2000}. We measure the cross spectrum, instead of the halo-halo power spectrum, to reduce the amount of shot noise. The linear halo bias parameter is then given by the large scale limit of $b(k,M)$, i.e.,  $b(M) = b(k \ll 1,M)$. We only consider the halo mass bins for which $b(k,M)$ has clearly saturated to its asymptotic value on large scales.

The simulation results show that, within the errorbars, the linear halo bias parameter for the three models is indistinguishable at all epochs shown. This shows that the modifications to gravity in the $\nloc$ model are not strong enough to modify substantially the way that dark matter haloes trace the underlying density field. The ST formula, Eq.~(\ref{eq:st-halo-bias}), reproduces the simulations results very well. Note also that there is little difference between the curves computed using the fitted (solid lines) and the standard (dashed lines) $(q,p)$ parameters of Table \ref{table:qp}. We conclude the same as in the case of the mass function that, in the context of the $\nloc$ model, there is no clear need to recalibrate the $(q,p)$ parameters in order to reproduce the bias results from the simulations.
 
\subsection{Halo concentration}\label{sec:results-c}

\begin{figure*}
	\centering
	\includegraphics[scale=0.39]{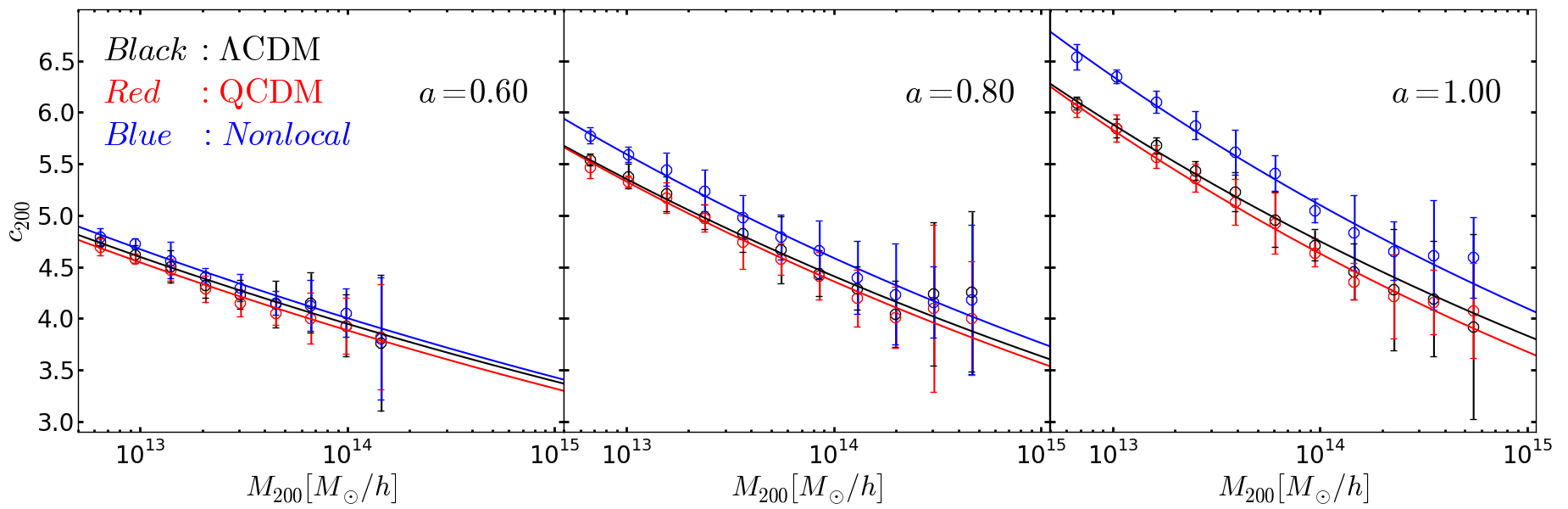}
	\caption{Halo concentration-mass relation in the $\lcdm$ (black), $\qcdm$ (red) and $\nloc$ (blue) models, for three epochs $a = 0.6$, $a = 0.8$ and $a = 1.0$, as labelled. The symbols show the mean halo concentration in each mass bin, considering only haloes (and not subhaloes) with mass $M_{200} > 1000\times M_p \sim 5\times 10^{12} \msun$, where $M_p = \rho_{m0}L^3/N_p$ is the particle mass. In the $a = 0.6$ panel, we omit the results from the two highest mass bins due to their few number of objects. The errorbars show twice the variance of the mass-binned mean concentration across the five realizations of the initial conditions. The solid lines show the best-fitting power law relations of Table \ref{table:ab}.}.
\label{fig:c}\end{figure*}

\begin{table}
\caption{Concentration-mass relation best-fitting $(\alpha, \beta)$ parameters in the parametrization ${\rm log}_{10}(c_{200}) = \alpha + \beta{\rm log}_{10}\left(M_{200} / \left[10^{12} M_{\odot}/h\right]\right)$ to the simulation results at $a = 0.6$, $a = 0.8$ and $a = 1.0$. The uncertainty in the values of $\alpha$ and $\beta$ is $\Delta_{\alpha} = \Delta_{\beta} = 0.001$. These are the parameters that minimize the quantity $\sum_i (c_{200}^{\rm sims}(M_i) - c_{200}^{\rm param}(M_i, \alpha, \beta) )^2/(2\Delta c_{200}^{\rm sims}(M_i))^2$, where $c_{200}^{\rm sims}(M_i)$ is the mean halo concentration measured from the simulations, $\Delta c_{200}^{\rm sims}(M_i)$ is the variance of the mean across the five realizations and $c_{200}^{\rm param}(M_i, \alpha, \beta)$ is the concentration given by the parametrization. Here, the index $i$ runs over the number of mass bins.}
\begin{tabular}{@{}lccccccccccc}
\hline\hline
\\
Model  & \ \ $a = 0.6$ & $a = 0.8$ & $a = 1.0$&\ \ 
\\
              & \ \ $(\alpha,\beta)$ & $(\alpha,\beta)$ & $(\alpha,\beta)$&\ \ 
\\
\hline
\\
$\lcdm$                    &\ \  $(0.729 , -0.066)$ &  \ \ $(0.813 , -0.084)$ & $(0.863 , -0.093)$&\ \ 
\\
$\qcdm$                  &\ \  $(0.726 , -0.068)$ &  \ \ $(0.814 , -0.087)$ & $( 0.866 , -0.100)$&\ \ 
\\
$\nloc$                     &\ \  $(0.737 , -0.067)$ &  \ \ $(0.834 , -0.086)$ & $(0.898 , -0.095)$&\ \ 
\\
\hline
\hline
\end{tabular}
\label{table:ab}
\end{table}

Figure \ref{fig:c} shows the halo concentration-mass relation for the $\lcdm$ (black), $\qcdm$ (red) and $\nloc$ (blue) models, at $a = 0.60$, $a = 0.80$ and $a = 1.00$. The symbols correspond to the mean values of $c_{200}$ identified in the same halo catalogues used in Fig.~\ref{fig:mf}. For all models, and at all epochs and mass scales shown, one sees that the halo concentrations are well fitted by the power law function (solid lines), 

\bq
{\rm log}_{10}(c_{200}) = \alpha + \beta{\rm log}_{10}\left(M_{200} / \left[10^{12} M_{\odot}/h\right]\right),
\eq
with the best-fitting $(\alpha, \beta)$ parameters given in Table \ref{table:ab}. In the $\nloc$ and $\qcdm$ models, one recovers the standard $\lcdm$ result that halo concentration grows with time at fixed mass, and that, at a given epoch, the concentration decreases with halo mass.

At early times ($a \lesssim 0.6$), all models predict essentially the same concentration-mass relation. At later times, however, the halo concentrations in the $\nloc$ model become increasingly larger compared to $\lcdm$. In particular, at $a = 1.00$, the halos are $\approx 8\%$ more concentrated in the $\nloc$ model, compared to $\lcdm$, for the entire mass range probed by the simulations. This can be attributed to a combination of two effects. Firstly, the enhanced structure formation in the $\nloc$ model may cause the haloes to form at earlier times. This leads to higher concentrations since the haloes form at an epoch when the matter density in the Universe was higher. Secondly, the increasingly larger value of $G_{\rm eff}$ is also expected to play a role via its effect in the deepening of the gravitational potentials. In other words, even after the halo has formed, the fact that gravity keeps getting stronger with time may also help to enhance the concentration of the haloes (see also Refs.~\cite{rebel1, Hellwing2010, Hellwing2010b, Hellwing2013a}). In the case of the $\qcdm$ model, one finds that the halo concentrations are hardly distinguishable (within errorbars) from those in the $\lcdm$ model, at all times and for all mass scales. This suggests that the differences between the expansion history of the $\qcdm$ and $\lcdm$ models (cf.~Fig.~\ref{fig:lingf}) are not large enough to have an impact on the formation time of the haloes. Once the haloes have formed in these two models, one can think as if the clustering inside these haloes decouples from the expansion. As a result, and since the gravitational strength is the same (cf.~Table \ref{table:models}), one sees no significant differences in the concentration of the haloes from the $\qcdm$ and $\lcdm$ simulations.

\subsection{Nonlinear matter power spectrum}\label{sec:results-pk}

\begin{figure*}
	\centering
	\includegraphics[scale=0.39]{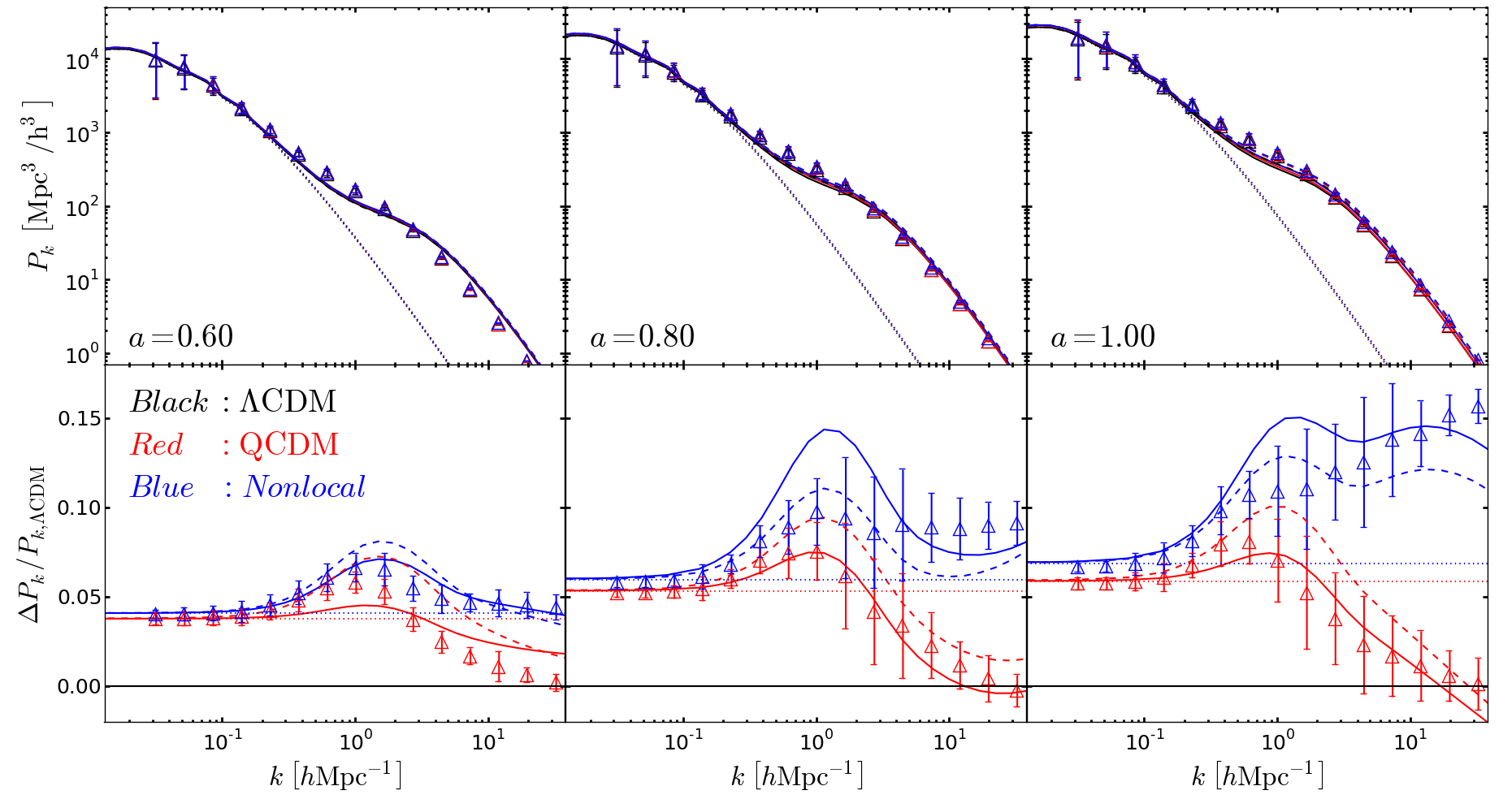}
	\caption{The nonlinear matter power spectrum (upper panels) in the $\lcdm$ (black), $\qcdm$ (red) and $\nloc$ (blue) models, at three epochs $a = 0.6$, $a = 0.8$ and $a = 1.0$, as labelled. The lower panels show the different w.r.t. $\lcdm$. The symbols show the simulation results, where the errorbars show twice the variance across the five realizations of the initial conditions. The solid lines show the halo model prediction obtained using Eq.~(\ref{eq:halo-model}), with the best-fitting $(q,p)$ parameters listed in Table \ref{table:qp}. The dashed lines show the power spectrum when using the standard ST $(q,p) = (0.75, 0.30)$ parameter values. The dotted lines show the result from linear perturbation theory. These lines are indistinguishable in the upper panels.}
\label{fig:pk}\end{figure*}

Figure~\ref{fig:pk} shows our results for the nonlinear matter power spectrum. The power spectrum from the simulations was measured using the {\tt POWMES} code \cite{Colombi:2008dw}. The solid (dashed) lines show the halo model prediction obtained using Eq.~(\ref{eq:halo-model}) with the fitted (standard) $(q,p)$ parameters of Table \ref{table:qp}. The dotted lines show the predictions obtained using linear theory. Next, we discuss these results separately for large, intermediate and small length scales.

{\it Large scales.} On scales $k \lesssim 0.1 h/\rm{Mpc}$,  the halo model is dominated by the 2-halo term, which is practically indistinguishable from the linear matter power spectrum. This is because, in the limit in which $k \rightarrow 0$, one has that

\bq
I(k) \sim \int {\rm d}M \frac{1}{\bar{\rho}_{m0}}\frac{{\rm d}n(M)}{{\rm dln}M} b_{\rm lin}(M) = 1,
\eq
where we have used the fact that $|u(k \rightarrow 0, M)| \rightarrow 1$ and  the last equality holds by the definition of the ST mass function and halo bias \cite{Cooray:2002dia}. In fact, in the standard halo model approach, replacing $P_k^{2h}$ by $P_{k, \rm lin}$ in Eq.~(\ref{eq:halo-model}) leads to practically no difference. As a result, the agreement between the halo model and the simulation results on large (linear) scales is always guaranteed.

{\it Intermediate scales.} On scales $0.1 h/{\rm Mpc} \lesssim k \lesssim 1h/{\rm Mpc}$, the halo model underpredicts slightly the power spectrum measured from the simulations, for all models and at all epochs shown. This is due to a fundamental limitation of the halo model on these scales, which follows from some simplifying assumptions about the modelling of halo bias on these intermediate scales (see e.g.~Sec.~IV.F of Ref.~\cite{2014JCAP...04..029B} for a simple explanation). In fact, the so-called {\it halofit} model arises as an alternative to the halo model that is more accurate on these intermediate scales \cite{Smith:2002dz, Takahashi:2012em, Zhao:2013dza}.  Nevertheless, in terms of the relative difference to $\lcdm$, the halo model limitations cancel to some extent, which leads to a better agreement with the simulation results. Focusing on $a = 1.00$, the halo model predictions for the $\qcdm$ model reproduce very well the results from the simulations. The predictions for the $\nloc$ model, although not as accurate as in $\qcdm$, still provide a fair estimate of the enhancement of the clustering power relative to $\lcdm$. For example, at $k \approx 1 h/{\rm Mpc}$ and $a = 1.00$, the simulations show an increase of $\approx 11\%$ in the power relative to $\lcdm$, whereas the halo model predicts an enhancement of $\approx 15\%$, which is similar. Finally, it is worth mentioning that the performance of the halo model when ones uses the standard $(q,p) = (0.75, 0.30)$ values (dashed lines) is comparable to the case where one uses the values that best fit the mass function results (solid lines).

{\it Small scales.} On scales of $k \gtrsim 1h/{\rm Mpc}$, the halo model predictions are dominated by the 1-halo term, whose agreement with the simulations becomes better than on intermediate scales, especially at $a = 1.00$. There are still some visible discrepancies at $a = 0.60$, which are similar to those found in Ref.~\cite{2014JCAP...04..029B} for Galileon gravity models. These discrepancies are however likely to be related with some of the assumptions made in the halo model approach, namely that all matter in the Universe lies within bound structures, which is not true in the simulations. However, similarly to what happens on intermediate scales, the halo model performs much better when one looks at the relative difference w.r.t. $\lcdm$. The predictions obtained by using the standard $(q,p)$ parameter values (dashed lines), although not as accurate as the results obtained by using the fitted $(q,p)$ values (solid lines), are still able to provide a good estimate of the effects of the modifications to gravity in the $\nloc$ model on the small-scale clustering power.
\bq
\nonumber
\eq

In the $\qcdm$ model, the relative difference w.r.t. $\lcdm$ becomes smaller with increasing $k$. In particular, for $k \gtrsim 10 h/{\rm Mpc}$ at $a = 1.0$, the clustering amplitude of these two models becomes practically indistinguishable. This result can be understood with the aid of the halo model expression for the 1-halo term, $P_k^{1h}$, (cf.~Eq.~(\ref{eq:halo-model-terms})), which depends on the halo mass function and concentration-mass relation. Firstly, one notes that for smaller length scales, the integral in $P_k^{1h}$ becomes increasingly dominated by the lower mass end of the mass function. Consequently, the fact that the mass function of the $\qcdm$ model approaches that of $\lcdm$ at low masses (becoming even smaller for $M \lesssim 5\times 10^{12} \msun$ at $a = 1.00$), helps to explain why the values of $\Delta P_k/P_{k, \lcdm}$ decrease for $k \gtrsim 1h/{\rm Mpc}$. Secondly, according to Fig.~\ref{fig:c}, the halo concentrations are practically the same in the $\lcdm$ and $\qcdm$ models. In other words, this means that inside small haloes (those relevant for small scales), matter is almost equally clustered in these two models, which helps to explain why $\Delta P_k/P_{k, \lcdm}$ is compatible with zero for $k \gtrsim 10h/{\rm Mpc}$ (Ref.~\cite{Brax:2014yla} finds similar results for $k$-mouflage gravity models). 

The same reasoning also holds for the $\nloc$ model, which is why one can also note a peak in $\Delta P_k/P_{k, \lcdm}$ at $k \sim 1h/\rm{Mpc}$. However, in the case of the $\nloc$ model, the mass function is larger at the low-mass end and the halo concentrations are also higher, compared to $\qcdm$ and $\lcdm$. These two facts explain why $\Delta P_k/P_{k, \lcdm}$ does not decrease in the $\nloc$ model, being roughly constant at $a = 1.00$ for $k \gtrsim 1h/\rm{Mpc}$. In particular, we have explicitly checked that if one computes the halo model predictions of the $\nloc$ model, but using the concentration-mass relation of $\lcdm$, then one fails to reproduce the values of $\Delta P_k/P_{k, \lcdm}$ on small scales. This shows that a good performance of the halo model on small scales is subject to a proper modelling of halo concentration, which can only be accurately determined in N-body simulations.

\subsection{Nonlinear velocity divergence power spectrum}\label{sec:results-ptt}

\begin{figure*}
	\centering
	\includegraphics[scale=0.39]{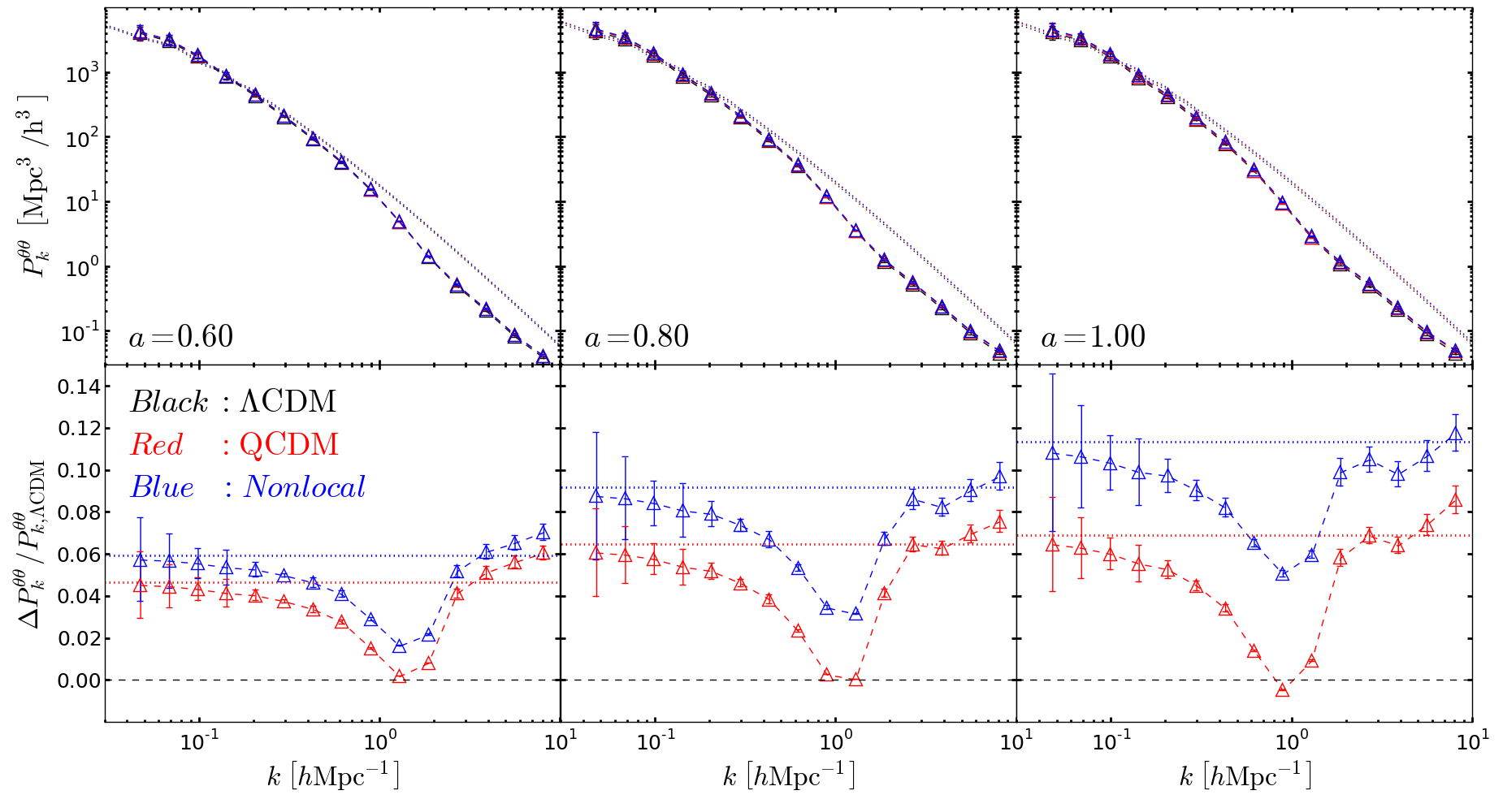}
	\caption{The nonlinear peculiar velocity divergence power spectrum (upper panels) in the $\lcdm$ (black), $\qcdm$ (red) and $\nloc$ (blue) models, for three epochs $a = 0.6$, $a = 0.8$ and $a = 1.0$, as labelled. The lower panels show the difference w.r.t. $\lcdm$. The symbols show the simulation results, where the errorbars show the variance across the five realizations of the initial conditions. The dashed lines only link the symbols to help the visualization. The dotted lines in the bottom panels show the prediction of linear perturbation theory.}
\label{fig:ptt}\end{figure*}

Figure \ref{fig:ptt} shows the nonlinear velocity divergence power spectrum, $P_{\theta\theta}$,\footnote{Here, $\theta$ is the Fourier mode of the divergence of the peculiar physical velocity field $v$, defined as $\theta(\vec{x}) = \nabla v(\vec{x})/H_0$.} for the three models of Table \ref{table:models} and for $a = 0.60$, $a = 0.80$ and $a = 1.00$. The computation was done by first building a Delaunay tessellation using the particle distribution of the simulations \cite{cv2011, sv2000}, and then interpolating the density and velocity information to a fixed grid to measure the power spectra. The upper panels show that on scales $k \lesssim 0.1 h/{\rm Mpc}$, the results from the simulations of all models approach the linear theory prediction, which is given by 

\bq\label{eq:pttlin}
P_{\theta\theta}^{\rm linear} = a^2\left(\frac{H}{H_0}\right)^2f^2P^{\rm linear}_{k},
\eq
where $P^{\rm linear}_{k}$ is the linear matter power spectrum and $f = {\rm dln}\delta_{\rm lin}/{\rm dln}a$. On smaller scales, the formation of nonlinear structures tends to slow down the coherent (curl-free) bulk flows that exist on larger scales. This leads to an overall suppression of the divergence of the velocity field compared to the linear theory result for scales $k \gtrsim 0.1 h/{\rm Mpc}$, as shown in the upper panels.

In the lower panels, the simulation results also agree with the linear theory prediction for $k \lesssim 0.1 h/{\rm Mpc}$. On these scales, the time evolution of the power spectrum of all models is scale independent and the relative difference encapsulates the modifications to the time evolution of $P^{\rm linear}_{k}$, $H$ and $f$, in Eq.~(\ref{eq:pttlin}). On smaller scales, the values of $\Delta P^{\theta\theta}_k/P^{\theta\theta}_{k, \lcdm}$ decay w.r.t.~the linear theory result until approximately $k = 1 h/{\rm Mpc}$. This suppression follows from the fact that the formation of nonlinear structures is enhanced in the $\qcdm$ and $\nloc$ models, relative to $\lcdm$ (cf.~Figs.~\ref{fig:mf} and \ref{fig:pk}). Hence, on these scales, the suppression in the velocity divergence caused by nonlinear structures is stronger in the $\qcdm$ and $\nloc$ model, compared to $\lcdm$. Finally, on scales $k \gtrsim 2-3 h/{\rm Mpc}$, the relative difference to $\lcdm$ grows back to values comparable to the linear theory prediction. On these scales, one does not expect haloes to contribute considerably to $P^{\theta\theta}_k$ for two main reasons. First, as haloes virialize, the motion of its particles tends to become more random, which helps to reduce the divergence of the velocity field there. Secondly, and perhaps more importantly, $P^{\theta\theta}_k$ is computed from a volume-weighted field, and as a result, since haloes occupy only a small fraction of the total volume, they are not expected to contribute significantly to the total velocity divergence power spectrum. On the other hand, considerable contributions may arise from higher-volume regions such as voids, walls or filaments, where coherent matter flows exist. For instance, matter can flow along the direction of dark matter filaments, or inside a large wall or void that is expanding (see e.g.~\cite{2014MNRAS.441.2923C, 2013MNRAS.429.1286C, 2014MNRAS.441.1974L}). These small scale flows are larger in the $\qcdm$ and $\nloc$ models at a fixed time, as shown by the growth of the values of $\Delta P^{\theta\theta}_k/P^{\theta\theta}_{k, \lcdm}$ on small scales. 

On scales $k \gtrsim 2-3h/{\rm Mpc}$, one may find it odd that the $\qcdm$ model predicts roughly the same matter power spectrum as $\lcdm$ (cf.~Fig.~\ref{fig:pk}), but has a different velocity divergence power spectrum. This has to do with the weight with which different structures contribute to $P_k$ and $P^{\theta\theta}_k$. For instance, $P_k$ is computed from a mass-weighted density field, and hence, it is dominated by the highest density peaks, which are due to dark matter haloes. In other words, it is very insensitive to the behavior of the clustering of matter in voids, walls or filaments due to their lower density. On the contrary, $P^{\theta\theta}_k$, which is computed from a volume-weighted field, is forcibly less sensitive to dark matter haloes due to their low volume fraction. The values of $P^{\theta\theta}_k$ are then mostly determined by the velocity field inside voids, walls and filaments. These structures are typically larger than haloes and therefore they are more sensitive to the background expansion of the Universe. Consequently, they are more likely to be affected by modifications to $H(a)$, compared to haloes which detach from the overall expansion sooner. This can then explain the differences in the sizes of the modifications to $P_k$ and $P^{\theta\theta}_k$ on small scales in the $\qcdm$ model, relative to $\lcdm$. To test this we have computed $P^{\theta\theta}_k$ by artificially setting $\theta(\vec{x}) = 0$ in regions where the density contrast exceeds $\delta = 50$. This should roughly exclude the contribution from haloes to the values of $P^{\theta\theta}_k$. We have found no visible difference w.r.t.~the results of Fig.~\ref{fig:ptt}, which shows that the small scale behavior of the velocity divergence is not affected by what happens inside dark matter haloes. We have performed the same calculation, but by setting $\theta(\vec{x}) = 0$ whenever $\delta < 0$, to exclude the contribution from voids. We have found that at $a = 1$, the relative difference of $\qcdm$ to $\lcdm$ at $k \sim 10 h/{\rm Mpc}$ drops from $\sim 9\%$ (as in Fig.~\ref{fig:ptt}) to $\sim 7\%$.  This seems to suggest that the dominant effect in the small scale behavior of $P_k^{\theta\theta}$ comes from walls and/or filaments. The velocity divergence in these structures is typically large (see e.g.~Fig.~2 of Ref.~\cite{2013MNRAS.428..743L}) and they also occupy a sizeable fraction of the total volume as well. A more detailed investigation of these results is beyond the scope of the present paper.

Focusing at $a = 1$, at $k \sim 10 h/{\rm Mpc}$ and relative to $\lcdm$, the velocity power spectrum in the $\nloc$ model is enhanced by $\sim 12\%$, and the matter power spectrum by $\sim 15\%$. On large (linear) scales the same figures are $\sim 12\%$ and $\sim 7\%$, respectively. The size of the modifications to the matter and velocity divergence power spectrum are rather similar, but the latter might be easier to measure as they are typically less sensitive to assumptions about baryonic processes such as galaxy bias. As an example, redshift space distortions (RSD) \cite{Scoccimarro:2004tg, 2011MNRAS.410.2081J, 2012MNRAS.427L..25J} are sensitive to the boost of the velocity field on large scales, and therefore can be used to test modified gravity models (see e.g.~Refs.~\cite{Wyman:2013jaa, 2012MNRAS.425.2128J}). The work of Refs.~\cite{2012PhRvL.109e1301L, Lam:2013kma} illustrated how the velocity distribution of infalling galaxies around massive clusters can be used to detect modifications to gravity (see also \cite{Zu:2013joa}). More recently, Ref.~\cite{Hellwing:2014nma} demonstrated that modified gravity models can leave particularly strong signatures in the velocity dispersion of pairs of galaxies on a broad range of distance scales. The level of precision of the data from future observational missions should prove sufficient to disentangle the differences depicted in Fig.~\ref{fig:ptt}. Such a forecast study would involve running simulations with better resolution and larger box sizes, and as such, we leave it for future work.

\section{Summary}\label{sec:summary}

We have studied the nonlinear regime of structure formation in nonlocal gravity cosmologies using N-body simulations, and also in the context of the semi-analytical ellipsoidal collapse and halo models. To the best of our knowledge, this is the first time the nonlinear growth of structure in nonlocal cosmologies has been studied. In particular, we investigated the impact that the modifications to gravity in nonlocal models have on the halo mass function, linear halo bias parameters, halo concentrations and on the statistics of the density and velocity fields of the dark matter.

The action or equations of motion of nonlocal gravity models are typically characterized by the inverse of the d'Alembertian operator acting on curvature tensors. Here, we focused on the model of Refs.~\cite{Maggiore:2014sia, Dirian:2014ara}, in which the standard Einstein-Hilbert action contains an extra term proportional to $\nloc$ (cf.~Eq.~(\ref{eq:action})). The constant of proportionality is fixed by the dark energy density today, and hence this model contains the same number of free parameters as $\lcdm$, although it has no $\lcdm$ limit for the background dynamics or gravitational interaction. 

Our goal was not to perform a detailed exploration of the cosmological parameter space in the $\nloc$ model. Instead, for the $\nloc$ model we used the same cosmological parameters as $\lcdm$ (cf.~Table \ref{table:params}). In this way one isolates the impact of the modifications to gravity from the impact of having different cosmological parameter values. Nevertheless, although a formal exploration of the parameter space in the $\nloc$ model is left for future work (see Ref.~\cite{swissgroup}), the comparison presented in Fig.~\ref{fig:cmb} suggests that the model fits the CMB temperature data as well as $\lcdm$. Our main results can be summarized as follows:

\bigskip

$\bullet$ The expansion rate in the $\nloc$ model is smaller than in $\lcdm$ at late times, and the gravitational strength is enhanced by a time-dependent factor (cf.~Fig.~\ref{fig:lingf}). Both effects help to boost the linear growth of structure (cf.~Fig.~\ref{fig:lingf}) and also speed up the collapse of spherical matter overdensities (cf.~Fig.~\ref{fig:dc}). In particular, at the present day, the amplitude of the linear matter (velocity divergence) power spectrum is enhanced by $\approx 7\%$ ($\approx 12\%$) in the $\nloc$ model, compared to $\lcdm$. These results are in agreement with Ref.~\cite{Dirian:2014ara}. The critical density for collapse today, $\delta_c(a = 1)$, is $\approx 3\%$ smaller in the $\nloc$ model, relative to $\lcdm$ (cf~Fig.~\ref{fig:dc}). For these results, the modified expansion history plays the dominant role in driving the differences w.r.t.~$\lcdm$, compared to the effect of the enhanced $G_{\rm eff}$.

\bigskip

$\bullet$ At late times ($a > 0.6$), the number density of haloes with masses $M \gtrsim 10^{12} \msun$ is higher in the $\nloc$ model, compared to $\lcdm$. The difference becomes more pronounced at the high-mass end of the mass function. In particular, at $a = 1$, haloes with mass $M \sim 10^{14} \msun$ are $\approx 10\%$ more abundant in the $\nloc$ model than in $\lcdm$. At $M = 10^{12} \msun$ this difference is only $\approx 2\%$. The effects of the modified $H(a)$ and $G_{\rm eff}$ on the enhancement of the high-mass end of the mass function are comparable.

The ST mass function describes well the absolute values of the halo number densities as well as the relative differences w.r.t.~$\lcdm$, for all of the epochs studied (cf.~Fig.~\ref{fig:mf}). We find that the use of the standard $(q,p) = (0.75, 0.30)$ ST parameter values provides a fair estimate of the modifications to the mass function in the $\nloc$ model. However, recalibrating these parameters to the simulation results helps to improve the accuracy of the fit (cf.~Table \ref{table:qp}).

\bigskip

$\bullet$ The linear halo bias parameter in the $\nloc$ model is barely distinguishable from that in $\lcdm$ for all masses and epochs studied (cf.~Fig.~\ref{fig:bias}). In other words, the modifications to gravity in the $\nloc$ model play a negligible role in the way dark matter haloes trace the underlying density field. The ST halo bias formula provides therefore a good description of the simulation results. There is also almost no difference between the semi-analytical predictions for the bias computed using the best-fitting and standard values for the $(q,p)$ ST parameters.

\bigskip

$\bullet$ The halo concentration-mass relation is well-fitted by a power law function (cf.~Fig.~\ref{fig:c}), but with fitting parameters that differ from those of $\lcdm$ (see Table \ref{table:ab}). For $a \lesssim 0.6$, the concentration of the haloes in the $\nloc$ model is roughly the same as in $\lcdm$, but it increases with time. In particular, at $a = 1.0$ ($a = 0.8$) and for all masses, haloes are $\approx 8\%$ ($\approx 4\%$) more concentrated in the $\nloc$ model, compared to $\lcdm$. This is likely to be mainly due to the enhanced $G_{\rm eff}$ on small scales, which helps to make the gravitational potential continuously deeper inside the haloes. On the other hand, the effects of the modifications to the expansion history in the $\nloc$ play a negligible role in changing the concentration of the haloes. This can be explained by the fact that the modifications to $H(a)$ are small enough not to have a significant impact on the formation time of the haloes. Consequently, once the haloes form, they detach from the expansion of the Universe and no longer "feel" the dynamics of the background.

\bigskip

$\bullet$ The modifications to gravity in the $\nloc$ model lead only to a modest enhancement of the clustering power. For instance, at $a = 1.0$ ($a = 0.8$) the amplitude of the nonlinear matter power spectrum is never larger than $\approx 15\%$ ($\approx 10\%$) on all scales (cf.~Fig.~\ref{fig:pk}). These differences might be hard to disentangle using data from galaxy clustering given the known uncertainties in modelling galaxy bias. On small scales, $k \gtrsim 1 h/{\rm Mpc}$, the differences w.r.t.~$\lcdm$ are completely determined by the enhanced $G_{\rm eff}$, and not by the modifications to $H(a)$.

At $a = 1.0$, the halo model describes the simulation results on large ($k \lesssim 0.1 h/{\rm Mpc}$) and small ($k \gtrsim 1 h/{\rm Mpc}$) scales very well. On intermediate scales and also at earlier times, the performance becomes worse due to known limitations of the halo model \cite{Cooray:2002dia, 2014JCAP...04..029B}. These follow from a number of approximations in the derivation of the halo model formalism, which sacrifice some accuracy in favour of analytical convenience. In terms of the relative difference w.r.t.~$\lcdm$, these limitations cancel out and the halo model describes the simulation results reasonably well for all epochs and scales. Moreover, the performance of the halo model formulae in describing the simulation results does not depend critically on the fitting of the $(q,p)$ parameters to the mass function. However, we have checked that the good performance of the halo model on small scales is subject to a correct modelling of the halo-concentration mass relation, which can only be properly determined via N-body simulation.

\bigskip

$\bullet$ Similarly to the case of the matter power spectrum, the modifications in the $\nloc$ model lead only to modest changes in the amplitude of the nonlinear velocity divergence power spectrum. In particular, at $a = 1.0$ ($a = 0.8$) the enhancement relative to $\lcdm$ is kept below $\approx 12\%$ ($\approx 10\%$) on all scales. However, measurements of RSD and/or galaxy infall dynamics are less subject to galaxy bias uncertainties, and therefore, might stand a better chance of distinguishing between these two models.

\bigskip

$\bullet$ The $\nloc$ model possesses no screening mechanism to suppress the modifications to gravity on small scales. As a result, Solar System tests of gravity can be used to constrain the model. For example, the $\nloc$ model predicts that $\dot{G}_{\rm eff}/{G} \approx 92 \times 10^{-13}\ \rm{yrs}^{-1}$, which is incompatible with the current bound from Lunar Laser Ranging experiments, $\dot{G}_{\rm eff}/{G} = \left(4 \pm 9 \right)\times 10^{-13}\ \rm{yr}^{-1}$ \cite{Williams:2004qba}. The local time variation of $G_{\rm eff}$ follows from the background evolution of the auxiliary scalar field $S$, and it seems nontrivial to devise a mechanism that can suppress it around massive objects or in high-density regions \cite{Babichev:2011iz, 2012PhRvD..85b4023K}. In this paper, we focused only on a particular choice of cosmological parameters. As a result, it might be possible that certain parameter combinations can be made compatible with Solar System tests, whilst still being able to yield viable cosmological solutions. Nevertheless, it seems clear that these tests should be taken into account in future constraint studies, as they might have the potential to rule out these models observationally.

\bigskip

In conclusion, the $\nloc$ model, although it has no $\lcdm$ limit for the dynamics of the background and gravitational interaction, exhibits changes of only a few percent in observables sensitive to the nonlinear growth of structure. Some of these effects are degenerate with baryonic mechanisms such as AGN feedback or galaxy bias, or even with massive neutrinos \cite{Baldi:2013iza, Shim:2014uta, Barreira:2014ija, Barreira:2014jha}. This makes it challenging to distinguish this model from $\lcdm$, but the precision of upcoming observational missions such as Euclid \cite{Laureijs:2011gra, Amendola:2012ys}, DESI \cite{Levi:2013gra} or LSST \cite{2012arXiv1211.0310L} should make this possible. From an observational point of view, however, future nonlinear studies are only warranted provided the model is able to fit successfully the CMB data from Planck (see Ref.~\cite{swissgroup}), in a way that is also compatible with the constraints from Lunar Laser Ranging experiments. The latter requirement might be hard to satisfy due to the absence of a screening mechanism.

\begin{acknowledgments}

We thank Philippe Brax, Marius Cautun, Yves Dirian,  Stefano Foffa, Ruth Gregory, Nima Khosravi, Michele Maggiore, Michele Mancarella, Alex Kehagias, Martin Kunz and Patrick Valageas for useful comments and discussions. We also thank Lydia Heck for invaluable numerical support. This work was supported by the Science and Technology Facilities Council [grant number ST/L00075X/1]. This work used the DiRAC Data Centric system at Durham University, operated by the Institute for Computational Cosmology on behalf of the STFC DiRAC HPC Facility (www.dirac.ac.uk). This equipment was funded by BIS National E-infrastructure capital grant ST/K00042X/1, STFC capital grant ST/H008519/1, and STFC DiRAC Operations grant ST/K003267/1 and Durham University. DiRAC is part of the National E-Infrastructure. AB is supported by FCT-Portugal through grant SFRH/BD/75791/2011. BL is supported by the Royal Astronomical Society and Durham University. WAH is supported by the ERC Advanced Investigator grant of C. S. Frenk, COSMIWAY, and by the Polish National Science Center through grant DEC-2011/01/D/ST9/01960. The research leading to these results has received funding from the European Research Council under the European Union's Seventh Framework Programme (FP/2007-2013) / ERC Grant NuMass Agreement n. [617143]. This work has been partially supported by the European Union FP7  ITN INVISIBLES (Marie Curie Actions, PITN- GA-2011- 289442) and STFC. 

\end{acknowledgments}

\bibliography{nonlocal-sims.bib}

\end{document}